\documentclass[twocolappendix,appendixfloats]{emulateapj}

\usepackage{float}
\usepackage{graphicx}
\usepackage{epstopdf}
\usepackage{amsmath}
\usepackage{apjfonts}
\usepackage{nicefrac}
\usepackage{hhline}
\usepackage{tcolorbox}
\tcbuselibrary{breakable}
\tcbuselibrary{skins}

\usepackage[colorlinks,citecolor=cyan,urlcolor=blue]{hyperref}

\renewcommand{\abstractname}{\footnotesize ABSTRACT}

\shorttitle{\small{SETI {\it in vivo:} testing the we-are-them hypothesis}}
\shortauthors{\small{Makukov M.A. and shCherbak V.I.}}

\begin{document}

\title{SETI \textit{in vivo}: Testing the We-Are-Them Hypothesis}

\author{Maxim A. Makukov\altaffilmark{1$\ast$} {\normalfont and} Vladimir I. shCherbak\altaffilmark{2}}

\altaffiltext{1}{Fesenkov Astrophysical Institute, Almaty, Republic of Kazakhstan}
\altaffiltext{2}{Al-Farabi Kazakh National University, Almaty, Republic of Kazakhstan}
\altaffiltext{$\ast$}{Corresponding author. Email: \href{mailto:makukov@aphi.kz}{\texttt{makukov@aphi.kz}}}

\begin{abstract}
\begin{tcolorbox}[colback=gray!8!,boxrule=0.35pt,before upper={\parindent10pt}]	
After it was proposed that life on Earth might descend from seeding by an earlier extraterrestrial civilization motivated to secure and spread life, some authors noted that this alternative offers a testable implication: microbial seeds could be intentionally supplied with a durable signature that might be found in extant organisms. In particular, it was suggested that the optimal location for such an artifact is the genetic code, as the least evolving part of cells. However, as the mainstream view goes, this scenario is too speculative and cannot be meaningfully tested because encoding/decoding a signature within the genetic code is something ill-defined, so any retrieval attempt is doomed to guesswork. Here we refresh the seeded-Earth hypothesis in light of recent observations, and discuss the motivation for inserting a signature. We then show that ``biological SETI'' involves even weaker assumptions than traditional SETI and admits a well-defined methodological framework. After assessing the possibility in terms of molecular and evolutionary biology, we formalize the approach and, adopting the standard guideline of SETI that encoding/decoding should follow from first principles and be convention-free, develop a universal retrieval strategy. Applied to the canonical genetic code, it reveals a nontrivial precision structure of interlocked logical and numerical attributes of systematic character (previously we found these heuristically). To assess this result in view of the initial assumption, we perform statistical, comparison, interdependence, and semiotic analyses. Statistical analysis reveals no causal connection of the result to evolutionary models of the genetic code, interdependence analysis precludes overinterpretation, and comparison analysis shows that known variations of the code lack any precision-logic structures, in agreement with these variations being post-LUCA (i.e. post-seeding) evolutionary deviations from the canonical code. Finally, semiotic analysis shows that not only the found attributes are consistent with the initial assumption, but that they make perfect sense from SETI perspective, as they ultimately maintain some of the most universal codes of culture.
\end{tcolorbox}
\end{abstract}

\section{Background}
\label{sec:background}
\addtocounter{footnote}{-3}

While much skepticism remains about practical feasibility of interstellar colonization as a low-risk endeavor, populating extrasolar habitats remotely with microorganisms is far less challenging and feasible, with certain efficiency, even with today's terrestrial technologies. Directed cosmic seeding has already been proposed as a mid-term project for humanity, with the aim to secure and spread terrestrial life-form (\citealt{Meot-Ner1979,Mautner2004,Tepfer2008,Gros2016,Sleator2017}). With no any benefits expected, such project stems from \textit{valuing} life and thus implies biotic ethics as its sole motivation (\citealt{Mautner2009}).

As potentially habitable planets much older than Earth are being discovered (\citealt{Anglada-Escude2014,Campante2015}; see also \citealt{Lineweaver2001}), it appears possible that terrestrial life might as well descend from seeding by an earlier civilization, as first suggested in \citet{Haldane1954} and \citet{Shklovskii1966}, and considered in more detail by \citet{Crick1973} and \citet{Crick1981}. In fact, if abiogenesis is uncommon, the seeded-Earth hypothesis becomes even more likely, as follows from a simple statistical argument. Abiogenesis could have happened when first habitable planets formed, and eventually a first intelligent species evolved. Having concluded, based upon available evidence, that abiogenesis is unique, this civilization would take on spreading life by cosmic seeding; on some of the seeded planets with conditions suitable for evolution of complex life (\citealt{Ward2003}), second-generation intelligent species might have evolved (and at least some of them might repeat the cycle of seeding). Thus, even if abiogenesis happened only once, there might have existed more than one civilization in the Galaxy (and even more planets inhabited with simple life-form). Hence, with the course of time it becomes more probable for an intelligent observer to find herself on a planet where life descends from seeding rather than on a planet where abiogenesis took place. This conclusion is particularly hard to avoid  under the view that ``the real feat of evolution is the origin of the cell~-- the rest is history'' (\citealt{Koonin2011b}), i.e. that abiogenesis is unique or very rare but that life, once appeared on a planet with favorable conditions, might evolve to complex forms relatively easily (e.g., \citealt{ConwayMorris2011,Bains2016}).

The seeded-Earth hypothesis has been criticized for not solving the problem of life's origin but ``merely shifting it to another place'', which is entirely irrelevant as it was never supposed to tackle that problem. Difficulties in understanding how life emerged from non-living matter are not, {\it per se}, the motivation for the seeded-Earth hypothesis. It is motivated by the suit of observations, which include ancient exoplanets, early appearance of life on Earth, and universality of certain cellular mechanisms (most notably of the genetic code), strongly suggesting that all terrestrial life derives from a small group of identical cells, or even perhaps from a single cell (\citealt{Doolittle2000,Koonin2011b}). As for invoking Occam's razor to eliminate the extra assumption of interstellar transfer, it would be justified only if conditions for life's emergence were identical in all planetary systems ever existed. The latter, however, is by itself an extra assumption.

Another common objection is that it is hardly possible to successfully deliver microbes to an exoplanet many light years away. Indeed, to this end, one would need an automated navigation technology with decision-making capabilities next to artificial intelligence (\citealt{Gros2016}). However, things might be much more prosaic if seeding is targeted not at individual exoplanets, but at star-forming regions that collapse to produce star clusters with hundreds and even thousands of stars (and their planetary systems). The clusters are disrupted, for a number of reasons, typically within just a few million years (\citealt{deGrijs2010}), and stars with their planets are then dispersed in the galactic disc. Apart from that no artificial intelligence is needed to seed a huge collapsing molecular clump, this option provides two extra bonuses. First, efficiency of seeding might be enhanced, as among thousands planets to be formed at least some might turn out to be suitable for the seeds to adapt and trigger biological evolution. Second, interference with indigenous life in this case is practically excluded. The downside is that the seeds have to additionally survive through the process of planetary formation, but this might be compensated with the swarm strategy (\citealt{Mautner1997}): it is not difficult to produce billions and billions of microbial seeds, and send them using a number of inexpensive probes (e.g., propelled by solar sails) headed into the same star-forming region.

As now established, most stars are formed in a clustered mode (\citealt{deGrijs2010}), and our Solar System is also believed to have formed in an open cluster (\citealt{Pfalzner2013,Kouwenhoven2016}). Thus, if terrestrial life derives from seeding, probably it goes back to seeding the original protocluster clump. If so, evidence that life inhabited Earth already at final stages of its formation (\citealt{Bell2015}) is expected rather than surprising -- in this case microbes are present around in cryptobiotic state when planets are being born, and, once within appropriate settings such as hydrothermal vents (\citealt{Weiss2016}), they resume metabolic activity and proliferate. Besides, with precision astrometry of the Gaia mission, there are attempts to identify the lost siblings of the Sun, i.e. the stars originated in the Sun's birth cluster (\citealt{MartinezBarbosa2016}), and the chances are these might host planets with our cosmic cousins, even if only in microbial form.

Would, however, an intelligent species decide to embark on cosmic seeding if natural panspermia can do that for free? Panspermia may indeed efficiently work within planetary systems, especially at early stages of their formation (\citealt{Mileikowsky2000}), and perhaps between planetary systems which are still within a star cluster (\citealt{Valtonen2009}). However, natural transfer of viable microbes between planets orbiting different field stars has been shown to be extremely unlikely (\citealt{Melosh2003,Valtonen2009}). At any rate, directed seeding is an automated rather than haphazard process, with the seeds maximally protected, so all things being equal, the efficacy of seeding is higher compared to panspermia.\footnote{Note that natural panspermia within planetary systems and within star clusters might be of help in directed seeding targeted at star-forming regions, as it subsequently provides additional mixing of material.}

Thus, observationally there is nothing implausible about the seeded-Earth hypothesis; in fact, compared with many topics discussed in the context of extraterrestrial intelligence, such as interstellar colonization, Galactic Club, postbiological evolution, etc., the seeding hypothesis appears rather unimaginative. Being skeptical about the scenarios above, it provides a realistic and yet perhaps the most fascinating resolution to Fermi's paradox, answering the question ``where are they?'' with ``they are here~-- we are all them'' (\citealt{Webb2015}).

\tcbset
{
	enhanced,
	oversize,
	left=1.5mm,
	right=1.5mm,
	coltitle=yellow!70!white!100!,
	colbacktitle=blue!5!black!52!,
	colback=gray!12!,
	fonttitle=\bfseries,
	drop fuzzy shadow southeast=gray!75!,
	boxrule=0.25pt,
	colframe=black,
}

\begin{tcolorbox}[title={\textcolor{black}{\normalfont Box 1.} Aliens become kins}]
	The seeded-Earth hypothesis proposed by \citet{Crick1973} appears to have a peculiar psychological effect. On the part of the academic community, it is manifested in that some authors regard this hypothesis as a tongue-in-cheek by its very intent (despite there is effectively nothing precluding this possibility from being the case). On the part of the general public, some people find it psychologically unappealing that the origin of life on Earth might have something to do with ``aliens'', especially when faced with the distorted form positing that life on Earth was ``created by aliens''. Leaving aside the question of whether prior credence of hypotheses should be affected by personal psychological attitudes, we remark that the seeded-Earth hypothesis should be no more discomforting than sharing common ancestry with all organisms here on Earth. It is not implied at all that terrestrial life was created from scratch for the purpose of seeding; in fact, the very point of ethically motivated seeding is to secure {\it existing native} life-form. In this regard, the very word ``aliens'' is something of a misnomer in this case, as the senders and the seeds' descendants share the same cellular ancestry.
\end{tcolorbox}

\section{A Testable implication}
\label{sec:prediction}

As the motivation behind seeding is ethical (valuing life), it is appropriate to consider other potential issues related to ethics. One is that seeding may interfere with indigenous life, but, as mentioned earlier, it is effectively avoided with targeting star-forming regions. There is another ethical issue. Delivering microbial seeds to other habitats, the senders hope (at least implicitly) that some of them might ultimately lead to evolution of an intelligent species. But if that happens~-- would it be ``good'' of the senders to leave those intelligent species without any clue about their descendant origin?

We make no illusion that this question is not debatable, and that the implication we have in mind is of absolute character; rather, it is a conditional prediction, with the condition that the senders were motivated to pass information to potential intelligent descendants by sending an intended artifact with the seeds. But we emphasize that this issue is in no way different from the one discussed in SETI/METI community: the motivation behind sending an artifact with the seeds is the same as behind any other interstellar messaging (by radio or with matter-packet artifacts). In most cases (e.g., Voyager Records) two-way communication, and hence gain of knowledge or other benefits, are hardly expected; the prime aim of the senders in these cases is to announce their own existence and thereby provide knowledge to other potential recipients. Thus, the motivation here is, again, ethical in its nature (for discussion of ethics and motivations in the context of SETI see, e.g., \citealt{Lemarchand2009,Vakoch2011,Vakoch2014}). In case of seeding, then, messaging does not introduce much of an extra assumption, because the working hypothesis here already implies ethical motivation behind the seeding itself. In fact, there is a stronger ethical background to messaging at seeding than to usual active SETI: it is on the responsibility of the senders to decide whether or not intelligent observers which might evolve from the seeds will be aware of their descendant origin. 

Assuming that the senders decide to pass at least minimal information to potential intelligent descendants, the questions is then how would they do that. An obvious option of an artifact attached to the probes that will carry the seeds is of no use~-- any passive material will disintegrate long before intelligent species might evolve. But even if it stays preserved, there is practically no chance it will be subsequently found after billions of years of evolution. In this situation the only available ``time capsules'' for an artifact are the seeds themselves~-- after all, life is a process, not a substance, and this process is stable and multiplying. An artifact or a ``signature'' designed to be inherited from the seeds and to endure through evolution is guaranteed to be delivered to intended recipients, as they will carry it with them all along. Besides, it is guaranteed to be delivered at the right moment in their cultural evolution, when they reach the point of scientific inquiry and technological level needed to look into cells. We will refer to this implication of the seeding hypothesis as biological SETI, or bioSETI for short. The question then comes to whether it is possible to embed an artifact into the seeds that will stay preserved while cells multiply and evolve.
\medskip

\begin{tcolorbox}[title ={\textcolor{black}{\normalfont Box 2.} bioSETI vs. Intelligent Design}]
	According to a common misconception, bioSETI is a form of Intelligent Design (ID) because both look for intelligent signatures in living cells; ID opponents and proponents alike succumb to this confusion. In fact, however, bioSETI is diametrically opposite in its premises~-- whereas ID ultimately seeks to argue {\it against} natural evolution, bioSETI attempts to answer how, {\it given} natural evolution, a durable signature might be encoded into evolving cells (correspondingly, methods also differ~-- similar to traditional SETI, bioSETI employs semiotic approaches, not ID-arguments like irreducible complexity). Hence, a positive result in bioSETI might count as a failure rather than success for ID, as it would imply validity of the premise which ID claims to be false.
\end{tcolorbox}

\begin{tcolorbox}[title ={\textcolor{black}{\normalfont Box 3.} Signatures differ},before upper={\parindent10pt}]
	\noindent Conceivably, a civilization might pursue cosmic seeding for reasons other than securing and spreading life -- e.g., for terraforming a planet for future colonization or performing a grand experiment. In these cases the senders also might be motivated to implant a signature into the seeds for technical reasons (e.g., for subsequent contamination control). However, there is no requirement here to make the signature universally intelligible~-- e.g., just a few arbitrary reassignments in the genetic code will suffice to discriminate cell lineages later on. In contrast, in ethically motivated seeding the signature is intended for recipients that might evolve from the seeds, so it must be discoverable and bare recognizable hallmarks of artificiality (we discuss what this might mean in Sec.~\ref{sec:semiotic}). Thus, by definition bioSETI makes sense only under ethically motivated seeding; detecting ``technical'' signatures in other types of seeding is hardly possible without comparing with original pre-seeding cells.
	
	Also note that presence of a signature (if there is any) within cells still does not imply that terrestrial life was ``created'' by the senders. Admittedly, inserting a signature into a cell implies this cell being genetically engineered to certain extent; still, there is a huge complexity gap between, e.g., rewriting the genome of an existing cell (leaving practically all of molecular machinery unchanged) and creating an entirely new life-form {\it ex novo}.
\end{tcolorbox}

\begin{tcolorbox}[title ={\textcolor{black}{\normalfont Box 4.} Too speculative? No more than good old SETI}]
	A typical response to bioSETI is that ``it is too speculative''. We remind that speculation, by definition, is pure reasoning detached from the possibility of being empirically addressed. The very point of the present paper, however, is that it exactly purports to approach the subject empirically. But even if compared at {\it a~priori} stage, bioSETI appears to involve, in fact, even weaker assumptions than traditional SETI. For that, compare basic coarse-grained assumptions behind both. The basic assumptions of traditional SETI are: (1)~primitive life had emerged independently on at least one other planet, (2)~a civilization had evolved from that simple life, (3)~that civilization is/was willing to announce its existence to potential recipients through artifacts (including electromagnetic signals). The basic assumptions of bioSETI are: (1)~terrestrial life descends from seeding by an earlier civilization, (2)~this civilization had evolved from a primitive life-form, (3)~this civilization was willing to announce its existence to potential recipients through artifacts (including a signature in the seeds). Some readers might reckon that the list for bioSETI should be longer and include the assumption that abiogenesis had occurred on another planet previously, but this would be erroneous. That abiogenesis had occurred {\it once} is not an assumption, but a fact (we know it because we exist). That it occurred on another planet is directly implied in the first assumption for bioSETI (which makes no sense if terrestrial life emerged on Earth). Thus, the number of basic assumptions is the same in bioSETI and traditional SETI, and assumptions (2)~and (3)~are equivalent in both cases. However, the assumption (1) of traditional SETI, that abiogenesis had occurred multiple times, is far from being weak. Meanwhile, bioSETI does not rely on it at all (as it goes with a single occurrence of abiogenesis) and replaces it with a weaker assumption of the seeded Earth (the potential of an already existing civilization to perform cosmic seeding is, at very least, not lower than the potential of life to emerge from scratch yet another time).
\end{tcolorbox}

\bigskip

\section{Location of a potential signature in the cell}
\label{sec:signaturelocation}
An artifact embedded into the seeds has no foresight to know which lineage, if any, will lead to intelligent species, so it should be inherited from the seeds by all lineages all the way up the (local) tree of life. This already provides two criteria to narrow down the search. First, it implies that ultimately the artifact must be encoded in genomes (directly or not), since there are no epigenetic mechanisms of information transmission as global and as reliable as genetic inheritance. Second, since inherited from a single origin, the artifact should be of universal character for all organisms, at least if expected to be intact. Genomic sequence itself is the most straightforward option for embedding a message at seeding; in fact, this possibility was realized immediately after \cite{Crick1973} proposed the seeded-Earth hypothesis (see, e.g., in \citealt{Hoch1997}), and was considered afterwards within the context of SETI (\citealt{Marx1979,Yokoo1979,Freitas1983,Davies2011a}). Meanwhile, encoding non-biological information within DNA {\it in vivo} has already been practiced here on Earth for other purposes (e.g., \citealt{Wong2003,Gibson2010}).\footnote{It is remarkable, though, that one of the very first demonstrations of storing messages in DNA, accomplished as early as in 1988, had an immediate reference to SETI (\citealt{Davis1996}; see also in \citealt{Gibbs2001}).} However, a biologically neutral genomic segment carrying a message is unlikely to survive intact for billions of years of evolution. Indeed, many genomes have already been sequenced and cross-compared, and no segments without biological function and identical throughout all (or even most of) terrestrial life have been found, at least thus far.\footnote{Ultraconserved DNA segments with so far unknown biological role do occur within distinct multicellular lineages (\citealt{Ahituv2007}). However, they are far from being universal and apparently had appeared during evolution in higher phyla.}

It appears that the only way to preserve an artifact during evolution is to make it hitchhiking functional genomic elements exposed to strong purifying selection. It is well known that genomic elements most conserved throughout all terrestrial life are the genes of translational machinery (\citealt{Isenbarger2008,Koonin2011b}), so if there is an artifact in the cells, most likely it is coupled to these genes. It cannot reside in their DNA sequence, as it practically has no room for non-biological information. However, an artifact might be made to hinge on their functionality which is preserved unaffected during evolution~-- the genetic code itself. Clearly, a signature cannot be embedded into amino acids or codons~-- these are just sets of physically predefined molecular structures. {\it Useful information might be encoded into how these sets are mapped to each other}, because the mapping of the code is not physically predefined (exactly this fact makes reassignment of codons possible, as now routinely practiced in labs). With that, the genetic code is the least evolving part of cells; while the underlying translational machinery has experienced evolutionary variations (occurring neutrally or under positive selection), the ultimate assignments between codons and amino acids have remained frozen under exceptionally strong negative selection since the time of the last universal common ancestor (LUCA, the original seeds in case of seeding). The few known post-LUCA variations of the code (\citealt{Knight2001a}) and alternative decoding mechanisms (\citealt{Baranov2015,Swart2016}) suggest that under certain circumstances codon assignments are evolvable to some extent. Nevertheless, the ubiquity of the canonical version of the code implies that this version has been absolutely conserved since LUCA.\footnote{To a reader without biological background we suggest to refer, e.g., to Appendix A in \cite{ShCherbak2013} for a more detailed elucidation of how it comes that the genetic code is amenable to artificial modification and yet is immutable during evolution.} This, together with the fact that the genetic code is the focal point in the cell, makes it an option of choice for embedding a ``notification'' signature at seeding, as first pointed out by \citet{Marx1979,Marx1986}.\footnote{George (Gy{\"o}rgy) Marx was a Hungarian particle physicist and science communicator, best known for his research in lepton physics. He was also interested in astrobiology and SETI, and served, after Frank Drake, as the chairman of the Bioastronomy Commission of the International Astronomical Union (currently Commission F3 -- Astrobiology).} 

\bigskip

\begin{tcolorbox}[title ={\textcolor{black}{\normalfont \hypertarget{box5}{Box 5.}} Towards methodological concerns},before upper={\parindent10pt},float,floatplacement=p]
	\label{box:scimethod}
	\small
	\noindent An actual attempt to approach bioSETI empirically and to try to check for a possible signature in the genetic code immediately encounters methodological concerns questioning the very possibility of a scientifically valid analysis within such approach. As this is one of the major points of criticism against bioSETI, here we address it in detail.
	
	First we point out the difference between traditional SETI and bioSETI. In traditional SETI there are good reasons to believe that the {\it form} of an artifact alone (be it an electromagnetic signal or an inscribed matter packet) would indicate artificiality, regardless of its content (in fact, even if the artifact is unintended). Things are subtler with bioSETI where artificiality cannot be deduced merely from the form (be it genetic code mapping or genomic segment) -- an actual decoding attempt must be made to decide. Ultimately, exactly this aspect is the source for most methodological concerns we hope to deal with.
	
	The typical objection here is that it is impossible to know in advance what an intelligent signature in the genetic code might look like, and, in fact, it is not even clear what it means to encode a signature into the mapping between amino acids and codons; therefore, all possibilities are open and the analysis then turns into ``numerology''. However, it is not an uncommon situation when the most nontrivial part of the research is exactly in developing a well-defined framework for data analysis, not so much in analysis itself. If a well-defined methodology for a certain issue does not suggest itself at first sight, this does not imply it cannot be found after a more careful consideration (of course, this tells nothing about whether application of this methodology will yield positive or negative result).
	
	In another version, the objection goes that {\it ``if one looks hard enough, it is possible to find patterns almost anywhere''}. It is difficult to disagree with this assertion as human brain is notorious for its tendency to perceive patterns even in random information (the phenomenon known as apophenia and pareidolia). But, again, according to this view, testing bioSETI is just that~-- hard looking for ``interesting'' patterns without any guidelines, and this is where the preconception sits. After all, any empirical science, when it comes to testing its predictions, is exactly about that~-- finding certain patterns in observational/experimental data, but no one pretends this implies crunching that data with all imaginable techniques to find something ``interesting''.
	
	Specifically, the two overlooked points here are the following. First, encoding/decoding useful information within a mapping between two unordered sets (such as the genetic code) might have a well-defined meaning, since, as shown in Sec.~\ref{sec:strategy}, the very notion of a ``message'' is formalized in terms of mathematical logic as a mapping between an alphabet and a set of natural numbers. The second point is that guidelines, and quite restrictive ones, {\it do} exist in bioSETI, and they are the same as in traditional SETI/METI: if an artifact is intended to be universally intelligible, its encoding/decoding {\it must} follow from first principles and be free of anything dependent on conventional schemes (in other words, it {\it must} avoid arbitrariness). Applying these guidelines in case of bioSETI (Sec.~\ref{sec:fromaatocodons}), it is possible to achieve a universal {\it a~priori} strategy for signature retrieval (universal and {\it a~priori} in the sense that it applies to any genetic code and does not depend on what actually will or will not be found in the end). This does not imply that, apart from this universal strategy, the retrieval process cannot involve (non-arbitrary) steps peculiar to specific situation (Sec.~\ref{sec:specificsteps}). Nor does it predict precisely how the signature might look (SETI is not about deterministic processes). Nevertheless, the universal strategy enormously reduces the possibilities and does not leave much wiggle room.
	
	Thus, {\it the point of testing bioSETI is not about searching for ``interesting'' or ``obvious'' patterns} in the genetic code, which is in fact ill-defined, for there is no measure of interestingness and obviousness. Instead, it is about patterns that might be encoded/decoded via a mapping between two unordered sets with a specific and well-defined method (Sec.~\ref{sec:strategy}).
\end{tcolorbox}

\section{Preliminary exploration of the possibility}
\label{sec:preliminaries}

Before trying to develop a methodology for signature retrieval, we perform preliminary analysis of the very possibility that the genetic code {\it might} store an intelligent signature, and investigate whether this possibility might be ruled out {\it a~priori} by existing data. Specifically, there are three immediate questions to address: (1) Is there enough room in the genetic code for an intelligent signature? (2) Does it contradict anything known about the genetic code? (3) Is embedding a signature into the genetic code technically feasible?

\subsection{Is there enough room in the genetic code?}
\label{sec:EnoughRoom}
The first question is whether there is enough room in the genetic code for a distinctly intelligent signature. This might be readily estimated. The number of all surjective mappings of $m = 64$ codons to $k = 22$ elements (20 amino acids and start/stop signals) is $N = k! S(m,k) \approx 2.35 \times 10^{85}$, where $S(m,k)$ is the Stirling number of the second kind (\citealt{Roberts2009}). The self-information of a particular mapping is therefore $\log_2{N} \approx 284$ bits. However, not all of these mappings are biologically plausible. As suggested by known variations of the code, there is a biochemical constraint on its structure in that tRNAs cannot discriminate between third-base pyrimidines within codons (\citealt{Ronneberg2000}). So the number of biologically available codons is only $m = 48$, and the number of codes is then reduced to $1.45 \times 10^{63}$. Among these, about one in a million is at least as efficient at minimizing errors as the canonical code (\citealt{Freeland1998a}), so the final number of biologically plausible codes is $1.45 \times 10^{57}$, and the self-information is now 190 bits. Thus, only a third part of the self-information of the code is reserved for biology, so even under biological constraints there is enough room for a signature. Besides, if properly projected, the signature might overlap with biological requirements to employ the capacity of the code more efficiently.

Though ${\sim} 200$ bits is not a particularly large volume to allow a full-fledged ``message'', it certainly suffices for a signature whose sole purpose is to provide indication of intelligent intervention, a kind of {\it ``to whom it may concern: we were here''} message.\footnote{This is analogous to ``beacons'' in traditional SETI: their sole purpose is also to reveal an artificial signature and, thereby, the existence of an intelligent species who created that signature.} Just for comparison~-- equivalent amount of bits allows to encode the first 35 prime numbers, or value of a mathematical constant like $\pi$ or $e$ with an accuracy of $10^{-60}$. In fact, this capacity is even comparable to the self-information of some Earth-made SETI-messages, such as the experimental 551-bit pictogram composed by Frank Drake (\citealt{Sagan1978}).

\medskip

\begin{tcolorbox}[title ={\textcolor{black}{\normalfont Box 6.} Too small for a signature?}]
	A skeptic might contend that the genetic code is too small to draw definite conclusions about a possible artifact within its mapping. However, this assertion is somewhat biased, because the same amount of information (the mapping of the canonical code) does not seem to raise analogous issue within conventional approaches. E.g., hardly anyone calls into question the fact that the canonical mapping is highly error-robust, even though the data behind this conclusion takes up only about 7\% of the informational capacity of the code.
\end{tcolorbox}

\bigskip

\subsection{Is it compatible with the existing data?}
\label{sec:Compatible}

The second question is whether it makes sense to approach the genetic code with bioSETI in view of the existing data. There are several models (with adaptive, biosynthetic and stereochemical being the predominant ones, see in Appendix~\ref{sec:appendixStat}) that attempt to provide rationale for why the canonical genetic code employs the particular assignments between codons and amino acids that it employs; reviews of these models might be found, e.g., in \citet{Knight1999a}, \citet{Koonin2009}, and \citet{Koonin2017}. Still, the bottom line is that while there are several speculated mechanisms that could shape the mapping of the code, there is no data that provides a decisive answer ({\it ibid.}). Likewise, there is no data excluding the possibility that the mapping of the canonical code reflects artificial intervention for inserting a signature in case of seeding. That said, a biologist might point out that there are two things about the canonical mapping which, without reference to any model, argue in favor of its natural origin: its block structure and robustness to errors. Undoubtedly, these features render biological efficiency to the code: the first one optimizes the energetics of the decoding process as it reduces the number of tRNAs in use (and, in fact, might follow from biochemical constraint, see \citealt{Ronneberg2000}), while both of them contribute to minimization of deleterious effects of translation errors. But exactly for these reasons it pays to preserve both these features if the mapping is to be altered artificially for inserting a signature, so their occurrence in the terrestrial code is not at odds with bioSETI.

Other alleged evidences for certain pathways of the genetic code evolution are heavily model-dependent and, in some cases, are mutually contradictory. Whatever the case, they are not in conflict with bioSETI for the simple reason that the latter does not negate evolution of the code; it only implies that it evolved earlier and elsewhere, and then was used for embedding a signature in seeding. Thus, \citet{Fournier2007} deduce an imprint of a primitive genetic code in ancient protein lineages. This inference, however, is model-dependent: what they identified in the first place is the difference in amino acid usage in conserved positions of ancient (ribosomal and ATPase) and more recent proteins, and it is not obvious that it unambiguously reflects the expansion history of the code. But even if it does, this does not rule out bioSETI, because seeds would possess ribosomal and {ATPase} proteins inherited ultimately from ancient cells evolved elsewhere under primitive genetic code, whereas insertion of a signature would not modify these proteins. Curiously, some amino acids (in particular, Ile and Tyr) identified by \citet{Fournier2007} as late additions to the code, must have entered the code even in pre-translational era of its development, according to \citet{Rodin2011}. In fact, there is a wider conflict between data underlying the stereochemical and biosynthetic models: almost all amino acids with reportedly strongest affinities to their (anti)codons (\citealt{Yarus2009}) are found as products, rather than precursors, in the biosynthetic model of the code expansion (\citealt{Wong1975,DiGiulio2004}). The latter, in its turn, has been criticized for the lack of statistical robustness (\citealt{Amirnovin1997,Ronneberg2000}). Meanwhile, recent discoveries of manifestly versatile and extraordinarily adaptive properties of the canonical amino acid set signify that most, if not all, amino acids must have entered the code due to natural selection rather than due to affinities for trinucleotides or due to coevolution of biosynthetic pathways (\citealt{Philip2011,Ilardo2015}). With that, these findings do not preclude a seeding-signature whatsoever, because its insertion would only modify the mapping, without altering the amino acid alphabet itself (see below). The fact that few amino acids suggest, notwithstanding, a stereochemical era for the genetic code (\citealt{Yarus2009,Rodin2011}) does not rule out bioSETI as well, since insertion of a signature might have not affected some of the original assignments, leaving a stereochemical ``imprint'' from the primordial code (that might be even desirable at projecting a signature, as it would simplify its embedding {\it in vivo}). The conclusion is that while the existing data reveal certain degree of mutual conflict, practically none of them contradicts bioSETI.

Another possible objection to bioSETI is that, as discovered during last decades, there is much more to the genetic code than just a mapping~-- e.g., there are context-dependent decoding (\citealt{Brar2016,Swart2016}), effect of the code's redundancy on expression, folding, splicing and other non-translational processes (e.g., \citealt{Plotkin2011,Maraia2014}), participation of molecular machinery of the genetic code in non-translational processes in higher organisms (\citealt{Giege2008,Guo2013}), etc., so it appears preposterous to think that there might be an artifact in the code. However, these not-just-mapping features have evolved {\it after} LUCA (i.e. after seeding, if that is the case), and represent adaptations of certain mechanisms to the particular mapping of the genetic code, not the other way round (at least because the code is universal while these mechanisms are not).

\subsection{Is it technically feasible?}
\label{sec:Feasible}
The third question is whether it is possible to reassign codons globally (as would be obviously required for inserting a signature). The very possibility of reprogramming the genetic code has already been thoroughly demonstrated (\citealt{Liu2010,Chin2014}). Though there might exist certain technical challenges (\citealt{Lajoie2016}), with advanced methods of synthetic biology there seem to be no fundamental barriers to global reassignment of codons. First, at {\it in silico} stage, a new mapping of the code is projected (without altering the amino acid repertoire) in such a way that it both conforms to functional requirements and harbors an intelligent signature. Then, at {\it in vivo} stage, this mapping is introduced into cells (by tweaking the existing machinery, primarily \mbox{tRNAs}, aminoacyl-tRNA-synthetases, and release factors). The major challenge to radical reassignment of the code is that it interferes with codon usage which, in turn, affects a variety of non-translational mechanisms (\citealt{Makukov2014}). This makes global reassignment of the code hardly possible to be performed in a single step. Still, as demonstrated by \citet{Lajoie2013a} and \citet{Ostrov2016}, stepwise genome-wide recoding of genes is feasible, at least in prokaryotes. Thus, there is no reason why global modification of the code is impossible if codon reassignments are introduced sequentially, with periods of artificial selection in between to readapt the cells to new codon usage.

\medskip

\begin{tcolorbox}[title={\textcolor{black}{\normalfont Box 7.} The only opportunity}]
	To sum up, the genetic code possesses exactly the properties needed to store a durable signature in cells: it provides space for non-biological information and is amenable to artificial modification, but once fixed (by rewriting all genes with the new code) it remains practically immutable during evolution due to extremely strong purifying selection (correct translation of {\it all} genes hinges on it). No other thing in the cell provides such opportunity in seeding/bioSETI.
\end{tcolorbox}

\medskip

\section{Analysis methodology}
\label{sec:strategy}
If the genetic code does carry an intended artifact, what is next? To try to predict what a possible signature might represent and thus to define the strategy for its retrieval, it is instructive to reverse the situation and to envisage the task of embedding a signature into the code in the first place. So, at our disposal are the set of 64 codons on the one side, and the set of 20 canonical amino acids and two punctuation signs (start and stop) on the other side. Neither set is allowed to be modified (to avoid formidable task of redesigning the entire proteome and/or entire translational machinery), but the mapping between the two sets is editable, subject to functional requirements mentioned in Sec.~\ref{sec:Compatible}. The first impression suggests that encoding information into a {\it mapping} between two sets is something too exotic, but in fact it is not: {\it any} message is a tuple~-- an ordered sequence of symbols (letters/bits), and, in formal language, the latter is a (non-injective) mapping between a set of numbered positions (or simply natural numbers) and a set of symbols (an alphabet)~-- this is, indeed, the most general definition of a message (\citealt{Sukhotin1971}). With an alphabet alone there is no way to communicate; to produce information, symbols are arranged into ordered sequence, i.e. mapped to a set of numbered ``slots'' (spatial for written texts and temporal for speech and radio-encoded messages). The unusual aspect about employing the genetic code is not that it is a mapping, but that neither of its two sets is ordered/numbered {\it a priori}.

Thus, encoding information into a mapping between two unordered sets differs from usual messaging in only one aspect: one of the sets should be preliminarily ordered uniquely to provide numbered slots for the elements of the opposite set to be ``written'' against. Then, a mapping is projected such that, upon application to the ordered set, it produces a desired message in the opposite set. (In formal language, one of the sets is first mapped bijectively to the set of natural numbers of the same cardinality, and a message in this case represents the composition of two mappings.) Note that this method is general and applies to any pair of sets of unordered elements, so long as at least for one of the sets there exists a bijective and unique mapping to the set of natural numbers. We will refer to a message/signature encoded into a mapping between (initially unordered) sets as a codogram.

Accordingly, we now define the general retrieval strategy: {\it starting with either of the two sets (amino acids or codons), order its elements from first principles without reference to the assignments between the two sets; after that, apply the mapping of the canonical code to the ordered set, and analyze what it yields with elements in the opposite set}. Two points should be emphasized here. First, the signature (if there is any) is expected to be encoded in the mapping between the two sets, but this mapping is not referred to at the ordering stage, precluding interference with useful information. Second, ordering from first principles implies, in terms of SETI, that it should be observer-independent in the sense that it must not rely on conventional schemes and models but should use parameters and definitions on which all observers would agree.

As a first step, we note that the set of codons is uniform, as all codons are made up of the same four units (nucleotides). As a result, codons cannot be ordered in a unique way; they might be systematized only in a position-independent manner. In contrast, the set of amino acids comprises both a uniform part (standard blocks identical in all amino acids) and a non-uniform part (side-chains unique to each amino acid). In the end, each amino acid is unique, and it is straightforward to systematize them into ordered sequence using some of their quantifiable properties. Hence, we first try to go from amino acids to codons.

\bigskip

\section{From amino acids to codons: the universal strategy}
\label{sec:fromaatocodons}
\subsection{The optimal parameter}
\label{sec:optimalparameter}

The next question is then which one of many possible parameters that might characterize amino acid molecules should be chosen to order these molecules into a sequence. Though formalization above refers to natural numbers, many physical parameters are real-valued and might be used for ordering as well, so we will not restrict ourselves on purely formal ground. Still, within SETI approach there are two criteria, both following from first principles, which greatly restrict the choice and ultimately provide the answer.

The first criterion is that the parameter must not depend on conventional systems to be quantified, as well as on accuracy of its measurement (these two are interrelated since physical measurements always involve conventional systems of units). This is quite obvious: no one would expect all extraterrestrials to use the same systems of units and measurement methods of identical precision. An example of especially bad choice here is hydrophobicity, since no unique system for quantifying this property exists at all. Other poor examples include mass and volume of molecules, characteristic angles, pKa, etc. Dimensionless mass expressed, e.g., in units of $\nicefrac{1}{12}$ of $^{12}$C mass is also a poor option, since the very phrase ``expressed in units of something'' implies a prearranged convention. The conclusion is that, after all, the parameter must be countable rather than physically measurable.

The second criterion is that the parameter should not be degenerate~-- its values should be as unique to each amino acid as possible. We want to use this parameter to arrange amino acids into ordered sequence, so if it has many repeated values there will be uncertainties in placing amino acids (and, correspondingly, codons) relative to each other. The degeneracies of parameters might be compared using, e.g., Shannon entropy:
\begin{equation*}
H = -\sum_{i}P(x_i)\ln P(x_i).
\end{equation*}

\noindent Here $i$ runs over distinct values of a parameter $x$, and $P(x_i)$ is the frequency of the $i$-th value. There are twenty amino acids in the canonical code, so there are twenty possible values of a parameter. In one extreme, all twenty values are identical ($i=1, P_i=1$), and the entropy is $H_{min} = 0$ nats. In another extreme, all twenty values are unique ($i= 20, P_i = \nicefrac{1}{20}$), and the entropy is $H_{max} = 2.996$ nats. Table~\ref{table:CountableParams} shows values of five examples of countable parameters and the resulting entropies (as percentage of $H_{max}$). Among all countable parameters we had identified, the number of nucleons has the highest entropy. Therefore, with this parameter the ambiguity in placing the amino acids relative to each other is minimized.

\begin{deluxetable}{ccccccc}
	\tablecaption{The Shannon entropy for some of the countable amino acid parameters.}
	\tablehead{\colhead{Parameters $\to$} & \colhead{DB} & \colhead{H} & \colhead{A} & \colhead{Z} & \colhead{N}}
	\startdata
	Gly &  1 & 5 & 10 & 40 &  75 \\
	Ala &  1 & 7 & 13 & 48 &  89 \\
	Ser &  1 & 7 & 14 & 56 &  105\\
	Pro &  1 & 9 & 17 & 62 &  115\\
	Val &  1 & 11 & 19 & 64 &  117\\
	Thr &  1 & 9 & 17 & 64 &  119\\
	Cys &  1 & 7 & 14 & 64 &  121\\
	Leu &  1 & 13 & 22 & 72 &  131\\
	Ile &  1 & 13 & 22 & 72 &  131 \\
	Asn &  2 & 8 & 17 & 70 &  132\\
	Asp &  2 & 7 & 16 & 70 &  133\\
	Gln &  2 & 10 & 20 & 78 & 146\\
	Lys &  1 & 14 & 24 & 80 & 146\\
	Glu &  2 & 9 & 19 & 78 &  147\\
	Met & 1 & 11 & 20 & 80 &  149\\
	His & 3 & 9 & 20 & 82 &  155\\
	Phe &  4 & 11 & 23 & 88 &  165 \\
	Arg &  2 & 14 & 26 & 87 & 174\\
	Tyr & 4 & 11 & 24 & 96 & 181\\
	Trp & 5 & 12 & 27 & 108 & 204\\
	\hline \\
	Entropy (\%$H_{max}$): & 40.2\% & 67.6\% & 79.7\% & 85.2\% & {\bf 95.4\%}
	\enddata
	\tablecomments{DB~-- number of double bonds in a molecule; H~-- number of hydrogen atoms; A~-- number of all atoms in a molecule; Z~-- atomic number (number of protons); N~-- number of nucleons.}
	\label{table:CountableParams}
\end{deluxetable}

However, even with this parameter there still might be some uncertainty: the number of nucleons in a molecule depends on pH of the milieu, where some amino acids might be (de)protonated. Which pH value should be chosen? In SETI approach the answer is none. The first-principles choice here is to consider amino acids out of any context, since no extra information is needed in this case, whereas for various observers to agree on in-context molecules the context must be specified. The choice of isotopes also poses no problem. As far as we know, physics is the same everywhere in the Galaxy and beyond, so all observers would agree on what isotopes are most common. Therefore, to avoid ambiguity, for all parameters in Table~\ref{table:CountableParams} we consider only most common isotopes and out-of-context molecules.

\begin{figure*}
	\centerline{\includegraphics[width=17cm]{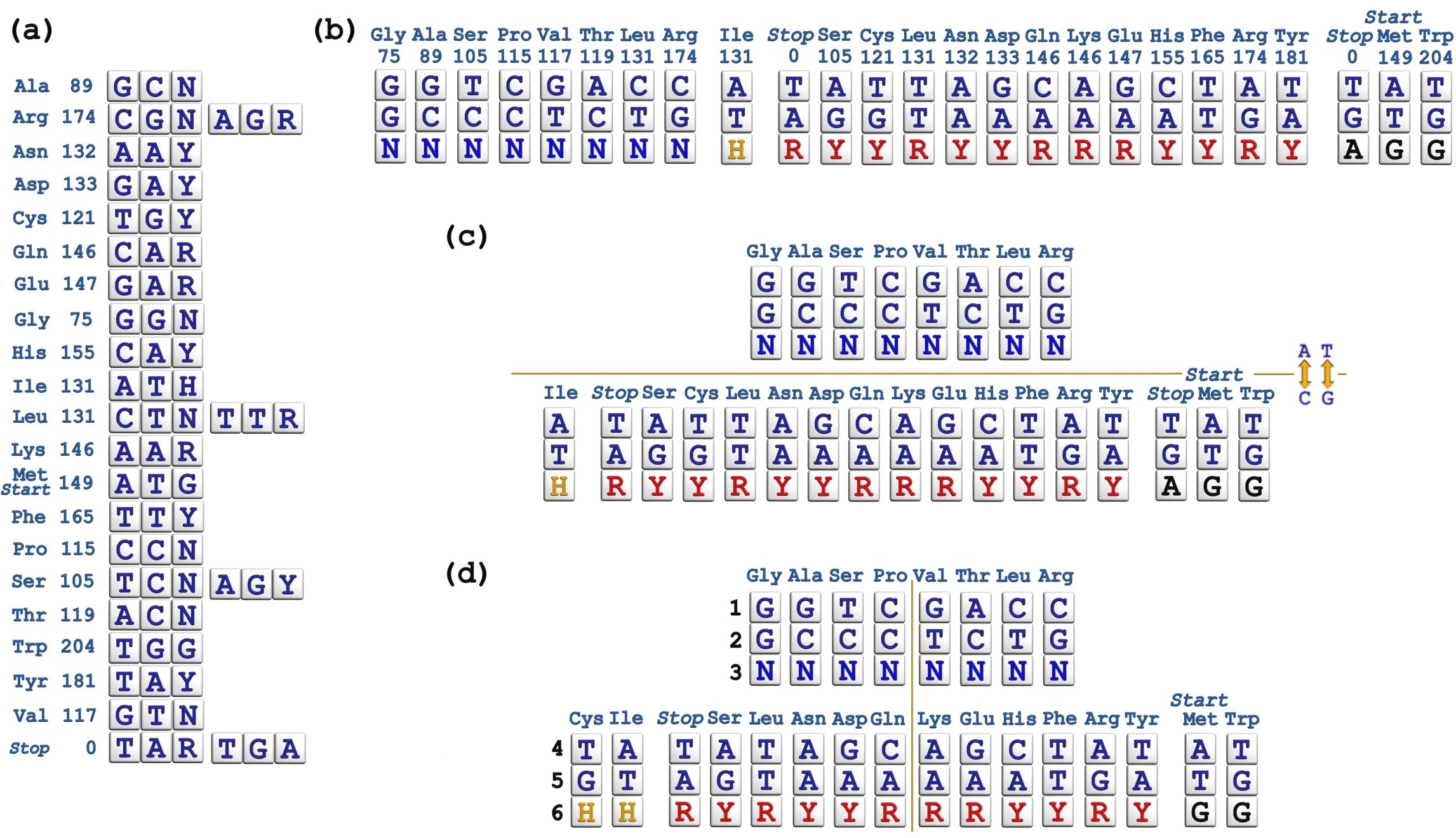}}
	\figcaption{Retrieving the codogram. {\bf (a)} Conventional inverse table of the canonical genetic code. Nucleon numbers of amino acids are shown. Codons are in the combined form (such combined representation of codons is occasionally used in the literature on the genetic code, see, e.g., \citealt{Knight1999a,Zhang2011}). {\bf (b)} Combined codons are put upright and arranged by nucleon numbers of their amino acids within distinct degeneracy groups. {\bf (c)}~Rumer's transformation is taken into account by opposing unsplit boxes (N-ending codons) to split ones. {\bf (d)} The final arrangement produced by symmetrization of the degeneracy pattern of the canonical mapping.\label{fig:codogram}}
\end{figure*}

\subsection{Combining the upright codons}
\label{sec:upright}
After identifying nucleon number as the optimal parameter, we employ it to arrange the amino acids into ordered sequence. The ordering stage is thereby completed, and we now apply the mapping of the canonical code to arrange corresponding codons. At this step we find that there are three codons in the canonical code that are not mapped to any amino acid and instead serve as stop-signals in translation (situation with start-signal is different as it is mapped to the same codon that encodes methionine). Since at this stage amino acids are already mapped to the set of nucleon numbers, to include stop-codons into systematization one has no choice but to assign zero nucleons to them.\footnote{A critically inclined reader may argue that the most natural thing is not to assign anything to stop-codons at all. But this is exactly what the symbol of zero is supposed to stand for~-- not assigning anything (in this case, in terms of amino acid nucleon numbers).}

We then also find that most amino acids are mapped to multiple codons, typically differing only in last nucleotide. E.g., CAT and CAC encode the same amino acid~-- which of them should come first? The only way to eliminate this ambiguity is to produce combined codons by applying conjoint symbols: N, standing for all of the four nucleotides; H, standing for T, C and A; Y for pyrimidines (T and C); R for purines (G and A). A biologist will recognize in these the standard nomenclature (\citealt{NC-IUB1985}) used in genomics and bioinformatics. With these new symbols, synonymous codons are combined into a single codon, e.g., CAT~+ CAC $\to$ CAY, GTT~+ GTC~+ GTA~+ GTG $\to$ GTN (see Fig.~\ref{fig:codogram}a).\footnote{In the canonical mapping three amino acids are encoded by six codons which cannot be combined into a single codon without over- or underrepresentation. E.g., leucine is encoded by CTT, CTC, CTA, CTG, TTA, TTG. These cannot be reduced to YTN, since that would include also TTT and TTC encoding phenylalanine. Reducing to YTR, we lose codons CTT and CTC. The only unambiguous representation of leucine therefore is with two combined codons CTN and TTR.} The difference from the standard application is that while in DNA/RNA sequences these symbols stand for {\it one} of the specified (e.g., R stands for either A or G), in application to the genetic code they stand for {\it all} of the specified (R stands for both A and G). There are two first-principles implications to this: (1) combined codons should be put upright side-by-side, rather than arranged linearly into one-dimensional sequence, and (2) combined codons should be arranged in distinct degeneracy groups.

The reason is that the additional symbols carry different amounts of information. While each of the standard four symbols carries two bits, Y and R carry one bit each, and N carries no information at all. Therefore, arrangement of combined codons in a linear sequence leads to mixing of symbols with inequivalent informational weights. Turning the codons upright and aligning them in this fashion, one produces three horizontal strings in which all symbols have equivalent weights (Fig.~\ref{fig:codogram}b). For the same reason, codons with informationally inequivalent third symbols are kept apart, so 4-, 3-, 2- and 1-degenerate codons are aligned within four distinct groups.\footnote{Note that this also removes ambiguity in positioning two combined codons of six-fold degenerate amino acids (see previous footnote), since these codons fall into different degeneracy groups.} The four groups themselves are arranged by their degeneracy number increasing in the direction opposite to that of nucleon numbers inside them. Such arrangement does not single out left or right, so it complies with the principle of being convention-free. (In another alternative the two parameters~-- nucleon and degeneracy numbers~-- ascend in the same direction, which thus becomes distinguished; we consider both options in statistical test below).
\bigskip

\section{Two retrieval steps specific to the canonical code}
\label{sec:specificsteps}
\subsection{Rumer's bisection}
\label{sec:Rumer}
The universal strategy (applicable to any genetic code) is now accomplished, so we turn to the result in Fig.~\ref{fig:codogram}b. We find that the entire set of 64 original codons is now neatly divided into two equal halves: 32 codons form the group of degeneracy 4 (combined codons ending with N), and the rest 32 codons form the other three groups. Moreover, it takes little effort to notice that these two halves reveal a simple transformation rule: for each codon in one half, there is a corresponding codon in the opposite half related to the first one by interchanging each T with G, and each A with C. This regularity is easily observed even from conventional codon table, which, for the canonical mapping, comprises equal numbers of split and unsplit codon boxes (codon box is a group of four codons that differ only in the third position; in unsplit boxes all four codons encode the same amino acid, and split boxes encode more than one amino acid). Exactly in this way it was first discovered by Yuri Rumer just after the code was cracked (\citealt{Rumer1966}).\footnote{English translation has been published recently, see \citet{Rumer2016}.} Moreover, while Rumer's transformation T$\leftrightarrow$G, A$\leftrightarrow$C interconnects globally the two halves of the code, the other two possible pair inversions (T$\leftrightarrow$C, A$\leftrightarrow$G and T$\leftrightarrow$A, G$\leftrightarrow$C) are found to be distributed equally between codons of the two halves. In the end, the overall arrangement of inversions in the canonical code is the most symmetric possible.\footnote{It might seem that such inversions occur whenever the code is divided into two equal halves, but this impression is misleading. The division must be specific to produce a global inversion between the two halves.} Interestingly, Rumer's observation was rediscovered several times (\citealt{Danckwerts1975,Wilhelm2004}); however, evolutionary models of the code have little to say about it, so it has been ignored for the most part. Yet, it represents an intrinsic feature of the canonical mapping, so we incorporate it into systematization by opposing the 4-degenerate group to the rest of codons (Fig.~\ref{fig:codogram}c).

\subsection{Symmetrization of the canonical mapping}
\label{sec:symmetrization}
The resulting representation in Fig.~\ref{fig:codogram}c is a two-part structure with each part comprising three (horizontal) strings of symbols. Even perfunctory analysis reveals symmetric features about this representation, e.g., mirror symmetry of complementary nucleotides in the first string. As might be easily noticed, the structure of the lower part is close to be symmetric as well. In fact, even in traditional tabular or circular representations of the code one might readily see that the canonical mapping is almost symmetric in its degeneracy pattern. There are only two codon boxes~-- those beginning with TG and AT~-- that are split unequally. The minimal action to restore the symmetry in degeneracy pattern is the reassignment of TGA-codon from stop-signal to cysteine (note that it keeps Rumer's inversion in place). With that, not only the overall structure of the arrangement (Fig.~\ref{fig:codogram}d) becomes symmetric, but it now reveals a non-trivial hierarchical structure of embedded attributes. For reasons described further on, this structure might be justifiably called a codogram.

By itself, the reassignment of TGA to cysteine is not arbitrary: it is the only single-step unique action that restores the symmetry in the degeneracy pattern of the canonical mapping.\footnote{The other single-step option is the reassignment of ATA, but this one is arbitrary: to restore the symmetry in degeneracy, ATA might be reassigned to {\it any} amino acid, except methionine.} But why this step should be taken at all certainly calls for justification, apart from the fact that it leads to the codogram. We will provide such justification in Sec.~\ref{sec:analysis}, when we will have the entire result at hand. In brief, this context-prompted step allows to increase effective informational capacity; the original (unsymmetrized) mapping, when analyzed below with the opposite strategy starting from the set of codons, reveals another completely different structure of attributes.

\bigskip

\section{Result: the codogram}
\label{sec:codogram}
So, the final representation in Fig.~\ref{fig:codogram}d (the codogram) reveals a hierarchical structure of embedded attributes all of which are related to various kinds of symmetry. Referring the reader to Video~\hyperlink{videos}{1} for detailed description  of these attributes (see links at the end of the paper), here we provide a summary, without any statements about their possible origin (these are left for later sections). At the bottom level is the Rumer's transformation, i.e. the inversion symmetry between the two parts of the codogram. All strings of the codogram reveal nontrivial symmetries at the level of nucleotide symbols (trivial only in the string 3 which carries no information). Another attribute is found at the level of triplets~-- the syntactic symmetry of punctuation signs (start and stop) within the triplet reading frame embedded in the string 5, reinforced by the ``crossword'' feature. Next, behind each combined codon there is an amino acid molecule. This amino acid background has uniform part (standard blocks) and non-uniform part (side-chains). Application of addition~-- the basic arithmetical operation~-- over these uniform and non-uniform parts reveals, for the lower half of the codogram, a precise nucleon balance~-- the quantitative symmetry. Remarkably, application of the same operation to the embedded syntactic structure reveals precise quantitative symmetry as well.\footnote{Given this multitude of embedded symmetries, it might seem that some of them are redundant, and that it might be not surprising to find many symmetries within a structure which has been previously symmetrized. However, as will be shown in Sec.~\ref{sec:algebraic}, this hierarchy of symmetries is nontrivial, so there is no overinterpretation.}
	
As follows from Table~\ref{table:CountableParams}, though the entropy of nucleon numbers is high, it is still 4.6\% less than the maximum: there are two pairs of amino acids with equal nucleon numbers. Remarkably, the effect of this percentage residue on the codogram is minimized. While Leu and Ile (both have 131 nucleons) fall into distinct degeneracy groups, Gln and Lys (both have 146 nucleons) fall into the same group. However, being interchanged, they preserve all symmetries of the string 5 and K/M-mirror symmetry of the string 4.\footnote{K and M are also from NC-IUB notation; K = \{G, T\}, M = \{A, C\}.}
\medskip

\begin{tcolorbox}[title={\textcolor{black}{\normalfont Box 8.} Nothing new invented}]
    With hindsight, it might be noted that the codogram-retrieving strategy involves few simple steps which reduce, in effect, to mere combination of data and methods known previously: starting with conventional inverse table of the genetic code, we apply NC-IUB nomenclature (\citealt{NC-IUB1985}), systematization of amino acids by nucleon numbers (\citealt{Hasegawa1980}), and Rumer's bisection (\citealt{Rumer1966}); the only introduced features are vertical placement of codons and symmetrization of the canonical mapping.
\end{tcolorbox}

\bigskip

\section{The opposite strategy: from codons to amino acids}
\label{sec:opposite}
Addition is a commutative operation, so for quantitative symmetries positions of codons are in fact irrelevant. This brings us back to the alternative strategy, in which one starts with the set of codons, rather than amino acids. As mentioned in Sec.~\ref{sec:strategy}, there is no unique way to lay out codons into ordered sequence~-- whichever parameter is chosen for that, it will be highly degenerate as all codons are made up of the same four nucleotides. Still, codons might be classified in a position-independent manner. After a little thought, one might convince himself that there are not many ways to sort codons uniquely from first principles, without regard to amino acid assignments (the requirement of the general strategy). In fact, one of these sortings was employed by George Gamow to propose one of the earliest models of the genetic code before it was cracked (see in \citealt{Hayes1998}). In this classification, codons are sorted according to their nucleotide composition, regardless of how nucleotides are allocated within codons.\footnote{Codons with identical composition make up 20 groups (the number of combinations of four distinct elements taken three at a time), and Gamow thought this could explain why there are 20 amino acids in the code. Though his model did not prove to be correct, the logic of Gamow's sorting holds as it does not rely on amino acid assignments at all.} Complemented with the same pair inversions that are found in Rumer's bisection, Gamow's sorting reveals, upon application of the canonical mapping and still employing nucleon numbers for amino acids, global and local quantitative symmetries of nucleon sums, all of which, furthermore, reveal homogeneous notation in decimal three-digit register (Video~\hyperlink{videos}{2}). The codogram-harboring symmetrized version of the code shares some of the local nucleon balances in Gamow's arrangement; however, the entirety of the balances is revealed only in the original unsymmetrized mapping.

Another first-principles classification of codons is logically complementary to Gamow's case: it sorts codons according to the type of nucleotides in certain position within codons, regardless of their overall composition (e.g., the set of codons might be divided according to whether the first nucleotide is purine or pyrimidine). Upon application of the canonical mapping, one finds no quantitative symmetries in any of these sortings. However, these sortings are equally applicable to combined codons, and, remarkably, taking the same combined codons that make up the codogram in Fig.~\ref{fig:codogram} and rearranging them according to these sortings one finds two more nucleon balances (Video~\hyperlink{videos}{2}).\footnote{These balances are shared by both code versions since in combined-codons representation they are algebraically the same (TGN box yields Cys~+ Trp for the symmetrized code, and Cys + 0 + Trp for the canonical code).}

Yet another straightforward classification might be performed with individual nucleotides that result from formal decomposition of all codons (Video~\hyperlink{videos}{2}). For the canonical mapping, it also reveals a global quantitative symmetry and decimalism, algebraically independent of the previous ones. Basically, these are all first-principles sortings that might be performed with codons, regardless of amino acid assignments. 

\medskip

\subsection{Activation key: constraint turned into advantage}
\label{sec:prolineskey}
The fact that all quantitative symmetries described in the previous section are revealed with a formal nucleon transfer in proline (see Video~\hyperlink{videos}{2}) probably stems from the constraint of dealing with the fixed set of amino acids at embedding the signature. The structure of proline differs from that of the rest of the canonical set: its side chain is enclosed to its standard block in a circular fashion.\footnote{Technically, proline is an amino acid with the secondary $\alpha$-amine (an imino acid in obsolete terminology).} Due to this peculiarity, proline has unique roles in proteins (e.g., it is found in turning points of polypeptide chains), so excluding it from proteome or replacing it with a standard-structure amino acid is practically impossible. Yet, this peculiarity also breaks the perfect uniformity of standard block nucleon numbers, since proline's standard block has one nucleon less compared to other amino acids in the canonical set.

This constraint, however, might be turned into advantage at embedding a signature. An intelligent observer easily sees that the set of amino acids comprises uniform and non-uniform parts, and that proline is the only element in this set that disturbs the uniformity of standard blocks. Therefore, all quantitative symmetries might be still embedded into the code in a way {\it as if} proline had regular standard block but lacked one nucleon in its side chain. Then, restoring full uniformity with the transfer of one nucleon from side chain to standard block in proline, an observer ``switches on'' all of the quantitative symmetries (including even those that occur only between side chain sums). The advantage here is that this step is {\it purely formal}, and so it acts like an ``activation key'' effectively preventing quantitative symmetries from being misconstrued with any natural cause, however exotic it might be. Certainly, such step of context-prompted decoding cannot be predicted {\it a priori}, but it is easily observed at applying the general strategy when first nucleon balances are revealed, and then it appears to work faultlessly in further analysis; meanwhile, not a single nucleon balance is found in the canonical code with the actual proline. Note also that whether the activation key is applied or not does not affect codogram attributes retrieved with the first strategy starting with amino acids. But, remarkably, applied there, it reveals homogeneous decimal notation of side-chain nucleon sum in the upper part of the codogram (where proline resides), in accord with the decimalism in the lower part (see Video~\hyperlink{videos}{2}).

\medskip

\begin{tcolorbox}[title ={\textcolor{black}{\normalfont Box 9.} Context-prompted decoding}]
	Following the subject of Box~\hyperlink{box5}{5} , another methodological concern might be raised about ``context-prompted'' steps in the retrieval process (symmetrization of the canonical mapping, Sec.~\ref{sec:symmetrization}, and proline's activation key, Sec.~\ref{sec:prolineskey}). As explained in these sections, most probably these steps follow from specific constraints. However, it should be noted that with electromagnetically encoded messages in traditional SETI/METI prompted steps are, in fact, inevitable, and that does not seem to cause methodological issues. For example, imagine that extraterrestrials have received the Arecibo message sent from the Earth. What they detect in the first place is the sequence of 1679 bits. The necessary (but insufficient) condition to correctly decode the message is to properly format the sequence. There are many ways to do so (including not only raster format, but also spiral and 3D format, etc.), but only one of them leads to the correct result when the format is a 2D-raster with 73 rows and 23 columns. Nothing obliges the recipients to prefer this option over others; but it is {\it prompted} by the fact that the number 1679 is semiprime and is uniquely decomposed into 23 and 73.
\end{tcolorbox}

\bigskip
\medskip

\section{Analysis of the result}
\label{sec:analysis}
The entire picture is now summarized as follows. With the genetic code, there are two possible directions for embedding (and, correspondingly, retrieving) a signature, one starting with amino acids and another starting with codons. Remarkably, the mapping of the universal canonical code is such that {\it both} directions reveal multiple nontrivial attributes, based on one and the same parameter (nucleon number). With that, these attributes do not represent a collection of unrelated peculiarities; instead, they are surprisingly systematic. Thus, quantitative symmetries of nucleon sums, as well as their decimalism, are found in both retrieval directions~-- in the codogram and in position-independent sortings of codons. Furthermore, each nucleon balance is accompanied by the same nucleotide pair inversions that appear in Rumer's bisection behind the codogram.

Apart from the general strategy, both retrieval directions involve a step prompted by the context during the general-strategy analysis. In case of going from amino acids to codons, this step is the reassignment of TGA to cysteine, prompted by restoring the symmetry in the degeneracy pattern of the canonical mapping; this single step enables the hierarchy of symmetries in the codogram. In case of going from codons to amino acids, this step is the nucleon transfer in proline, prompted by restoring the uniformity of standard block nucleon numbers; this single step enables the entirety of the quantitative symmetries and their decimalism in position-independent sortings of codons. The likely reason for the latter step, as discussed in the previous section, follows from the constraint in dealing with the fixed set of amino acids; the reason for the symmetrization step is that it increases informational capacity, in the following sense.

Suppose that the symmetrized version served as the universal code. Applying the first strategy one then could easily arrive at the codogram, without need for a context-prompted step. However, applying the second strategy starting with codons, one could reveal only some of the local quantitative symmetries in Gamow's sorting and none in the decomposed code. Conversely, having the canonical mapping alone, one reveals both the codogram with the first strategy (this time involving the context-prompted step) and the entirety of quantitative symmetries with the second strategy. Thus, the reason for the step of TGA reassignment is that it allows to encode more information within the same data: having a single mapping one might arrive at complete pictures in both retrieval directions. It is also remarkable that the codogram of the symmetrized code provides independent confirmation that TGA should stand for cysteine in it (see Video~\hyperlink{videos}{1}, clip 11).

Yet, the version of the code with TGA encoding cysteine is also found in nature in euplotid ciliates. Most probably, this is a coincidental fact, which is not unlikely given that TGA is one of the ``hot spots'' in the code which have been reassigned during evolution independently in various lineages to at least three amino acids (\citealt{Ivanova2014}). After all, one does not have to be aware of the existence of this version in nature to deduce it from the canonical code, as evidenced by the fact that the codogram of the symmetrized mapping, first described in \citet{Shcherbak1988} and \citet{Shcherbak1989}, had been found before the euplotid code was discovered (\citealt{Meyer1991}).\footnote{It is not excluded, though, that both versions of the code~-- the canonical and the symmetrized~-- might have been present among original seeds, and euplotid ciliates then might descend from the seeds with the symmetrized code. In view of this, it is interesting to note that genomes of these ciliates have unique features distinguishing them sharply from other eukaryotes (\citealt{Hoffman1995}). It is also relevant to note that some evolutionary biologists find it plausible that LUCA could be eukaryote-like, with prokaryotes emerging subsequently through reductive evolution (\citealt{Forterre1999,Doolittle2000,Glansdorff2008}). In case of seeding this would be a natural scenario, since eukaryotes are preferable for seeding as they might be better at adaptability and might facilitate evolution of complex organisms more easily than prokaryotes (both types could be used for seeding as well).}

\subsection{Statistical analysis}
\label{sec:statistical}
To explore the possibility that the described attributes might have arisen as a byproduct of certain evolutionary pathways (such as optimization via positive selection) we perform comprehensive statistical analysis. Unlike statistical test described in \citet{ShCherbak2013}, we now take into account not only functional (model-independent) aspects, but also model-dependent considerations suggested by the three major models of the genetic code origin and evolution. For the convenience of the general reader who is not supposed to be acquainted with specifics of these models, we put detailed description of the statistical analysis into Appendix~\ref{sec:appendixStat}. We focus there on the first strategy that goes from amino acids to codons. It should be emphasized that since the universal retrieval strategy applies to any genetic code, the question of statistical estimation of the attributes revealed in the canonical code is therefore well-defined. As follows from the statistical test, the symmetries of strings in the codogram cannot be ascribed to individual or combined effect of evolutionary and functional aspects.
\medskip

\subsection{Interdependence analysis}
\label{sec:algebraic}
To make sure that the entire system of the described attributes is nontrivial, i.e. that there is no overinterpretation of a lesser number of regularities, we need to show that all of the attributes are mutually independent (in algebraic sense). This might be verified even with direct inspection: for each attribute it is possible to find codon reassignments which do not affect this particular attribute but destroy the others. If some of the attributes were trivially interdependent, they would be always preserved or broken simultaneously.

Consider simple examples. If we interchange amino acids Lys and Asn between their codons (AAY and AAR), the string symmetries in the codogram in Fig.~\ref{fig:codogram}d, as well as the embedded symmetric structure of syntactic signs are preserved, but the nucleon balance behind the syntactic structure is broken. Likewise, if we reassign two pairs of codon blocks~-- TAR with AGY and TAY with AGR~-- the strings of the codogram will be again symmetric. However, there will be no embedded syntactic structure or other symmetries at the level of triplets.

Nontriviality of the attributes revealed with the opposite strategy is verified in a similar manner. E.g., interchanging amino acids between codons ATG and CTG preserves all nucleon balances in Gamow's sorting, but destroys balances in the decomposed and combined-codons representations. Interchanging between CGA and GAC preserves balances in Gamow's and decomposed sortings, but breaks all balances in the combined-codons representation, while interchanging between GGG and AAA preserves only the balance of the decomposed code.

Apart from safeguarding against overinterpretation, this analysis also demonstrates that the attributes in the genetic code are tightly interlocked, to the extent that a random codon reassignment breaks typically all or at least most of them. Of course, this is not unexpected when it comes to encoding a signature into the mapping between two small sets. Still, the degree to which the entirety of the attributes described in Videos~\hyperlink{videos}{1} and \hyperlink{videos}{2} are compacted into a single code version (one-codon difference between the two versions notwithstanding) is impressive. Developing such a signature amounts to solving a hierarchical combinatorial problem -- an exercise quite challenging even with advanced computational technologies at hand.

\bigskip

\subsection{Comparison analysis}
\label{sec:comparison}
Apart from the canonical (and symmetrized) version of the code, there are more than 20 currently known versions in some lineages of simple organisms, mitochondria and plastids.\footnote{See, e.g., \url{http://www.ncbi.nlm.nih.gov/Taxonomy/Utils/wprintgc.cgi}} It is a common consensus that these variants represent post-LUCA evolutionary deviations from the canonical mapping (\citealt{Knight2001a}), so they might serve as a ``control group'' in comparison test.

To this end, we performed full-treatment analysis of each of the known code variants under the framework of bioSETI, applying the retrieval strategy in both directions. In case of going from amino acids to codons, the universal strategy is the arrangement of combined codons by nucleon numbers of their amino acids within distinct degeneracy groups (as in Fig.~\ref{fig:codogram}b), whereas Rumer's bisection (Fig.~\ref{fig:codogram}c) and symmetrization (Fig.~\ref{fig:codogram}d) were specific to the canonical mapping. Some of the alternative code versions do not have equal numbers of whole and split boxes, so the retrieval procedure stops already at the stage corresponding to Fig.~\ref{fig:codogram}b. Nevertheless, we analyzed them with and without opposing whole groups to split ones, no matter if such opposing revealed nucleotide inversions or not. We also checked if those mappings were a single reassignment away from full symmetry in the degeneracy pattern, and if they were (as occurred in only two cases), we also checked the resulting symmetrized ``codograms''. With the opposite strategy (also applicable to any genetic code), we considered all combinations of subsets within first-principles sortings of codons (in Gamow, combined, and decomposed representations). With that, nucleon sums were calculated both with and without proline activation key applied (all known versions of the code use the same set of amino acids, so there were no other instances for context-prompted decoding). All in all, we have found that, even though these variations differ from the canonical mapping in only one or few reassignments, they lack most of the attributes found in the canonical code; with that, they do not reveal new systematic attributes of any type. At best, some variants retain the symmetries in the upper part of the codogram and/or one or two quantitative symmetries in combined representation left over from the canonical code. This result is in full agreement with bioSETI: intact signature might be expected only in the original code, whereas in subsequent evolutionary deviations it is damaged or erased completely.
\medskip

\begin{tcolorbox}[title ={\textcolor{black}{\normalfont Box 10.} bioSETI and the scientific method},before upper={\parindent10pt},float,floatplacement=t]
\noindent Another misconception about bioSETI might be summarized as follows:
\vskip3pt
\noindent {\em ``The problem with this approach is that it is predicated on the notion of having a hypothesis in search of data, however science is usually performed in the reverse order~-- data is generated and then hypotheses are made and tested.''}
\vskip3pt
\noindent Indeed, according to the standard scientific method, science begins with observations, not hypotheses. The problem with the comment above is that it stems from confusion. First, bioSETI is not a hypothesis proper~-- the hypothesis is that terrestrial life descends from seeding (\citealt{Crick1973}), and bioSETI is a {\it testable prediction} of that hypothesis; even though it is conditional, rather than absolute (see Sec.~\ref{sec:prediction}), it is still a prediction, not a standalone assumption. Second, the seeded-Earth hypothesis itself begins, as it should, with observations (see Sec.~\ref{sec:background}). (Keep in mind that different branches of science allow different amounts of initial observations; e.g., the amount of data generated in genomics is vastly greater than the amount of data available to researchers in astrobiology and SETI.)
	
Furthermore, this prediction is specific: not only it says that an intelligent signature might reside in the canonical genetic code, it also predicts that, if it does, in existing variations of the code the signature will be absent (or present only partially), because these variations deviated from the original code during evolution after first cells appeared on Earth. This specificity is exactly what is observed: all known versions of the code, when analyzed with the method developed under bioSETI framework, display few of the attributes retained from the canonical code, and do not reveal new systematic precision attributes of any kind. 
\end{tcolorbox}

\subsection{Semiotic analysis: artificiality and content}
\label{sec:semiotic}
We now turn to the analysis of the result from SETI perspective, to see how much sense it makes within the initial assumption.

First it should be discussed what a ``sign of artificiality'' might mean. In most general terms, it means something peculiar to culture as opposed to nature. The distinctive aspect about culture is that it is based on systems of codes, i.e. systems of relations between signs and things they signify (\citealt{Lotman1978}), with most notable codes being languages and writing systems. We cannot hope to review semiotics in any depth here, so we refer the reader to other sources~-- see, e.g., \citet{Deacon1997,Sebeok2001} and, in the context of SETI, \citet{Vakoch1998acta,Vakoch1998leo}. What should be mentioned here is that signs fall into three basic categories: indexes, which refer to their signified via causal connections (e.g., alarm calls in animal groups), icons, which refer to their signified via physical similarity (e.g., pictograms), and symbols, which have neither resemblance nor causal connection to the signified, but refer to it via interpretation (e.g., almost all words in any human language). While indexes widely occur in nature and in animal world, icons and particularly symbols are specific to cultures, so it is them that might serve as indicators of artificiality. All Earth-made interstellar messages (their list might be found, e.g., in \citealt{Zaitsev2012}) were constructed using iconic and/or symbolic codes.\footnote{In semiotic terms, the genetic code itself is an indexical code: there is a biochemical causality to why a particular amino acid is assigned to a particular codon (chemical causal chains might be modified to reprogram the genetic code, but the causality is still there). Meanwhile, Morse code is a symbolic code; there is no causality (other than convention) to why a particular combination of dots and dashes should stand for a particular letter.}

The vast majority of symbolic codes, including languages and writing systems, are almost entirely arbitrary in their conventions. However, cultural universals might and do exist. While some of them (e.g., assignment of color terms) are due to common neurobiological basis (\citealt{Deacon1997}), others might be related to more general cognitive principles. It is not an exaggeration to say that zero-based positional notation is one of the most universal codes of culture (\citealt{Dantzig1954,Kaplan1999,Gazale2000,Ifrah2000,Seife2000,Chrisomalis2010,Mazur2014}). Non-positional notations (e.g., Roman numerals) are arbitrary: to encode numbers in these systems one has to know multiple culture-specific conventions. Placeholder positional notations (which still do not employ zero as a number on its own, but use a placeholder), such as Babylonian and Mayan systems, already reveal a good deal of structuring; but they are still ambiguous and calculation-inefficient. Only in case of zero-based positional notation, one needs to know a single rule (place-value principle) and a single number (radix) to represent all other numbers. This makes such notation efficient for calculations (\citealt{Zhang1995}); treating zero as a standalone number and incorporating it into the positional notation had played the key role in the progress of human society (\citealt{Dantzig1954,Ifrah2000,Mazur2014}), not to mention that without zero-based positional notation advanced computing technologies would be impossible.

Given such efficiency and universality, it is likely that during technological evolution civilizations typically converge on using zero-based positional notation (though, of course, the choice of the radix is still arbitrary); already within human society several cultures converged independently on using positional numeral systems (\citealt{Kaplan1999,Seife2000}). With that, zero-based positional notation is still {\it a code of culture, not a phenomenon of nature}; nature is indifferent to how intelligent species decide to represent quantities, just like it is indifferent to how they assign words or other symbols to signify objects. In this sense, preferred notation system is a stronger sign of artificiality than, e.g., prime numbers, which are often mentioned in the context of SETI simply because no natural process capable of generating such numbers is known, not because they are a code of culture.

The most straightforward way to demonstrate zero-based positional system with a preferred radix to another observer is to produce quantities which have homogeneous notation (i.e. composed of identical digits, such as 555) in that particular system; in a system with any other radix notation of the same quantities will not be homogeneous (symbols used to denote digits are, of course, arbitrary, but homogeneous notation is unique in that it repeats one and the same symbol, whichever it is). This is exactly what is observed with the attributes in the genetic code. Precise equalities of nucleon sums (see Video~\hyperlink{videos}{2}) are there regardless of what numeral system is used; however, if these sums are written down with one and the same system~-- the decimal one~-- they systematically display homogeneous notation in three-digit register.

The reason why this system happens to be exactly base-10 might be quite unpretentious. Obviously, three-digit numbers with homogeneous notation in a system with radix $r$ are always multiples of $111_r$. Furthermore, as might be easily proved, if $r = 3n + 1$, $(n = 1, 2,...)$, then the number $111_r$ is factorized as $3 \times k$, where $k = 3n^2 + 3n + 1$. In particular, for the decimal system $r = 10$, $n = 3$, and $k = 37$. This makes decimal system most convenient for encoding into the genetic code, since standard blocks~-- the uniform part of the amino acid set~-- have 74 nucleons, which is a multiple of 37.

But could it be that homogeneous notation in the genetic code is nothing but a trivial consequence of 37-multiplicity (even though the reason why there should exist such multiplicity might be unknown)? The answer appears to be no. First, for a number to have homogeneous notation in three-digit decimal register, 37-multiplicity is a necessary but insufficient condition (the sufficient one is 111-multiplicity). Second, in quantitative symmetries that occur between side-chain sums (e.g., in Gamow's sorting), these sums are also decimally homogeneous, even though summands themselves are not multiples of 37. Third, it is here that the activation key plays its crucial role: physically, there are neither quantitative symmetries nor decimally homogeneous nucleon sums in the code. All of them are accessible to an intelligent observer through the prompted step of nucleon transfer in proline. Thus, the fact remains that the attributes in the genetic code systematically reveal decimalism, and yet this decimalism does not reduce to physical multiplicity. The conclusion is that this feature is purely notational, so we are dealing with the code of culture.

But even more than that, the symbol of zero occurs in the codogram in an entirely independent fashion~-- not implicitly as an integral part of the positional notation, but explicitly as a standalone number. Namely, all of the string symmetries in the lower part of the codogram in Fig.~\ref{fig:codogram}d emerge only if the combined stop-codon is placed in front of its group, where the zero should sit. Now, suppose that some natural process could distinguish between nucleon contents of amino acids (and even separate them according to Rumer's inversion of their codons). The maximum that such process might do is to sort out the amino acids into ordered sequences. Still, there is no zero here. Zero appears upon application of the mapping, where certain codons are found to be unmapped to any amino acid. In terms of nucleon numbers, an intelligent observer readily interprets that as zero quantity and puts corresponding codons at the beginning of the sequence; however, for natural processes stop-codons are nothing more than a collection of atoms combined into molecules in a manner that facilitates binding of release factors that trigger termination of the translation at the ribosome. Thus, zero here is a {\it symbol} in the proper sense of the word: all codogram symmetries are maintained only if stop-codons are correctly {\it interpreted} in terms of nucleon numbers as something that designates absence of an amino acid.\footnote{Similar to the case of proline, this does not imply that stop-signal had to be introduced into the code for the purpose of encoding the notion of zero; rather, being present in the preexisting code, stop-signal might have found its corresponding role at projecting the signature.}

Some readers might conclude that it is not surprising to find zero here after it had been introduced by the authors themselves in Sec.~\ref{sec:fromaatocodons}, through assigning it to combined stop-codon. To be clear: the key point here is not about {\it introducing} zero, it is about {\it placing} zero in its proper position. Surely, interpretation of stop-codon as zero is introduced by an observer, but {\it correctness} of this interpretation is confirmed when stop-codon is placed in front of its series, as only in this case the codogram happens to reveal symmetries in its strings. In our view, this fact alone is sufficient to seriously regard the codogram as a candidate bioSETI signature.

\begin{tcolorbox}[title ={\textcolor{black}{\normalfont Box 11.} On the argument from ignorance},float,floatplacement=b]
	A skeptical reader might put forward the following epistemological argument:
	\vskip3pt
	\noindent {\it ``Nonrandomness alone is not a proof of artificiality. Because natural selection is not a random process, the found ordered structure in the genetic code might represent an outcome of evolution (even though we might not know why it had been selected for)}''.
	\vskip3pt
	\noindent It should not be overlooked that this ordered structure was found within the approach developed {\it specifically} to test bioSETI; it was found neither accidentally nor within evolutionary models of the genetic code. While this by itself does not prove that the finding is indeed related to bioSETI, it makes referring to our conclusion as being based on the argument from ignorance irrelevant: bioSETI is {\it not} motivated by the lack of a generally accepted theory of the genetic code origin and evolution. At any rate, ultimately our conclusion follows neither from nonrandomness nor even from the inability to explain the finding with existing evolutionary models. It follows from semiotic analysis showing that the ordered structure in question reveals culture-specific features. (That said, we do, of course, take evolutionary models into account in statistical analysis).
\end{tcolorbox}

Yet another semiotic phenomenon is found within the triplet reading frame of the string 5 in the codogram (see Video~\hyperlink{videos}{1}, clip 09). This frame reveals syntactic symmetry of the two punctuation signs (stop and start), with a homogeneous codon in the center. Remarkably, this symmetry pertains {\it solely to the semantics} of the punctuation signs; physically, the initiation and termination of protein translation are unrelated processes triggered by entirely different molecular complexes. Exactly for the reason that the structure in Fig.~\ref{fig:codogram}d encodes the concepts of positional notation, of zero, and of the semantic symmetry, one might justifiably call it a codogram.

The same features that attest to the artificiality simultaneously represent the content of the signature. One of the first things one encounters in studying mathematics is the notion of numeral systems~-- to express any values (be it prime numbers, fundamental constants or any other quantities) one already must have a notation system at hand. This makes number notation systems of prime significance to any (technology-oriented) society. It is therefore not surprising that zero-based positional notation had been included as part of the content in most Earth-made interstellar messages (the Arecibo message, Voyager Records, Evpatoria messages, and others), typically in their opening sections. Meanwhile, zero, as an abstract symbol that signifies not an object but absence of an object, is one of the landmark symbols in cultural evolution; it was discovered by several human cultures independently (\citealt{Kaplan1999,Seife2000}; interestingly, in some of these cultures calendar counting even began with zero, rather than one). Today we are so accustomed to both the zero and positional notation, that the crucial role they played in the progress of human society is often forgotten. In fact, zero and positional notation are regarded by some scholars as two of the greatest achievements in the history of culture (\citealt{Dantzig1954,Zhang1995,Ifrah2000}). We thus find that not only the structure of attributes in the genetic code is consistent with the initial assumption, but that it makes perfect sense from SETI perspective, as it encodes some of the most prominent universals of culture: the zero-based positional notation, encoded in the direction from codons to amino acids, and the notion of zero as a number in its own right, encoded explicitly in the opposite direction from amino acids to codons. This appears sufficient to interpret the message as \textit{``to whom it may concern: we were here''}.
\bigskip
\bigskip

\begin{tcolorbox}[title={\textcolor{black}{\normalfont Box 12.} On the evolution of intelligence},float,floatplacement=t]
As another argument against bioSETI (in fact, traditional SETI as well), sometimes it is claimed that the odds of intelligent life evolving from microorganisms are so negligible that it would make no sense to bother about inserting a signature into the seeds (or, for that matter, searching for signals from space), and this argument is presented sometimes in a manner as if the question of intelligence evolution has been settled whereas in fact it is open. The whole issue here deals with two distinct questions. The first question, pertaining to astrophysics and planetary science, is about conditions necessary for evolution of complex life, and things seem to be more or less clear in this case. Apparently, such conditions must be very specific; this includes not only appropriate distance from the host star, but also appropriate chemical composition of a planet, presence of an appropriate moon stabilizing the planet's axis, perhaps plate tectonics, and more (\citealt{Ward2003}). The second question, pertaining to evolutionary biology, is about the odds of intelligent forms evolving from microbial life provided that all planetary conditions are conducive. In this case, the answer is currently uncertain. As a trend, those evolutionary biologists who adopt that contingency is the predominant factor in evolution while natural selection plays a minor role would typically argue that such odds are near zero, referring to Stephen Gould's metaphor of replaying the tape of life and having radically different outcomes each time (\citealt{Gould1989}). Conversely, those who view evolution as mostly adaptive rather than neutral would tend to conclude that the odds might be not so small after all. There is even an extreme view that not only the emergence of intelligent beings is almost inevitable, but that these beings almost certainly will be human-like, as argued most notably by \citet{ConwayMorris2003a,ConwayMorris2003b,ConwayMorris2011} (replay the tape of life repeatedly and the outcome will be almost the same each time). This claim is based on the observation of ubiquity of convergent evolution (at organismal, cellular and molecular levels), which is taken, in fact, as the strongest evidence for adaptation itself. In a less extreme form the view that evolution of intelligent species from a simpler life is not that unlikely is shared by some other evolutionary biologists, see, e.g., \citet{Dawkins2004} and \citet{Gould1985} (the latter might sound surprising, given that Gould was the major figure in emphasizing the role of contingency in evolution; in fact, however, he admitted intelligence as an attainable evolutionary niche, though he was uncertain about specific forms that might fill this niche). For convergent evolution in general and with regard to intelligence in particular, see \citet{ConwayMorris2003a,ConwayMorris2003b,Rospars2010,McGhee2011,Lobkovsky2012,Cirkovic2014}.
\end{tcolorbox}

\appendix

\section{Description of statistical test}
\label{sec:appendixStat}
\setlength{\textfloatsep}{2pt plus 1.0pt minus 2.0pt}
In this test we focus on the symmetries of codogram strings. The general strategy described in the main text is of universal character and therefore applies to any genetic code. The steps peculiar to the canonical mapping are the opposition of unsplit boxes to split ones (in the canonical code this yields Rumer's transformation), and symmetrization of the degeneracy pattern (Figs.~\ref{fig:codogram}c and ~\ref{fig:codogram}d). In general, therefore, in analyzing computer-generated codes one should check symmetries in the codograms of the universal type (as in Fig.~\ref{fig:codogram}b). However, such codograms are less likely to produce symmetries by chance as they have longer strings, so to avoid overestimation, along with such universal-type codograms, for each code we also check symmetries in codograms with whole groups opposed to split ones, but we do not require that such opposition should reveal any of the Rumer's-like transformations, or even equal numbers of codons. Besides, in our algorithm (see below) the codograms of computer-generated codes are necessarily symmetrized.
\vskip4pt
{\bf The scoring scheme.} Strings in alternative codograms are analyzed for mirror symmetries in combination with all types of nucleotide inversions, and among these the maximum score is used as the final value. The score $S$ for an individual string is introduced as $S = W \times (L - P - E) = W \times L_{eff}$. Here $L$ is the length of a string; $P = 2^n$ is the ambiguity penalty, with $n$ the number of codon pairs with the same nucleon number which, being interchanged, affect the symmetry (if there is some); $E$ is 0 if the length of a string is even and 1 if it is odd (in this case symmetries are analyzed relative to the central symbol, which itself does not contribute to symmetries); $L_{eff}$ is the effective length of a string (the part of $L$ which contributes to symmetries); $W$ is the Shannon weight~-- the Shannon entropy of a string normalized to the maximum entropy of a string of the same length. This coefficient takes into account the fact that, e.g., symmetries are less likely to occur by chance in a quaternary string (containing all four types of symbols) than in a binary string of the same length (containing only two types of symbols, or revealing a symmetry only in one of the binary codings R/Y, S/W, or K/M).

If a string does not reveal any symmetry, its score is set to zero. The maximum score that a string might have is $L$. The score for the entire codogram is the sum of the scores of its strings. For each alternative code, the codogram is built in four variants~-- with and without opposing unsplit boxes to split ones, and, in each of these, nucleon numbers are arranged both in the same and opposite direction to that in which degeneracy sets are ordered. Among the four variants, the maximum score is taken. Partially degenerate strings, inversions between the two parts of the codogram and symmetries at the level of triplets are not analyzed. In this scheme, the score of the codogram in Fig.~\ref{fig:codogram}d is $S_0 = 30.5$.

\vskip4pt
{\bf Generating alternative codes.} As mentioned in the main text, a biologically plausible genetic code should have block structure similar to that of the real code. Therefore, our algorithm for generating alternative codes starts with constructing their block structures. We do not require that they should be completely identical to that of the real code, but require them to be on average similar. In all known versions of the code all strong boxes (SSN) are unsplit, while all weak boxes (WWN) are split, which is believed to follow from thermodynamics of the decoding process (\citealt{Lagerkvist1978}). Therefore, our algorithm is as follows: probability for strong boxes to stay unsplit, as well as for weak boxes to be split, is taken to be 0.9, while for all other boxes probability to be split is 0.5. If a box is determined to be split, then it has probability 0.9 to be split into equal blocks (2:2 codon ratio) and 0.1 to be split in the ratio 3:1 (boxes with 1:3 ratio are not generated, since no known tRNA discriminates between third-base pyrimidines). For simplicity, we do not generate codon boxes that encode more than two amino acids (or an amino acid and termination signal; note that in this case the numbers of combined codons in 3- and 1-degenerate groups are equal, i.e. codograms are necessarily symmetrized). Since all canonical amino acids and stop-signal should be recruited at least once, codes with less than 21 blocks are discarded.

The next step is the assignment of blocks to amino acids. Its algorithm depends on the null hypothesis; we have tested three null hypotheses reflecting three major models of the code evolution, as well as combinations thereof.
\vskip4pt

\def\arraystretch{1.3}
\begin{table}[b]
	\begin{center}
		\caption{Dissociation constants (from \citealt{Yarus2009}) and corresponding probabilities for amino acids to be assigned to their ``cognate'' codons.}
		\begin{tabular}{ccc}
			\hhline{===}
			Amino acid & $K_d$, M & $p$  \\
			\hline
			Arg &  $1.0 \times 10^{-6}$ & 0.99 \\
			Trp &  $1.2 \times 10^{-6}$ & 0.98 \\
			His &  $1.2 \times 10^{-5}$ & 0.90 \\
			Tyr &  $2.3 \times 10^{-5}$ & 0.88 \\
			Phe &  $4.5 \times 10^{-5}$ & 0.86 \\
			Ile &  $5.0 \times 10^{-4}$ & 0.77 \\
			Leu &  $1.1 \times 10^{-3}$ & 0.75 \\
			Val &  $1.2 \times 10^{-2}$ & 0.66 \\
			Gln &  $2.0 \times 10^{-2}$ & 0.64  \\
			\hline
			\label{table:stereodata}
		\end{tabular}
	\end{center}
\end{table}

{\bf Model 1: Adaptive hypothesis.} According to the adaptive model, the assignments of the genetic code had been ultimately shaped by natural selection for error minimization. This model stems from the fact that the canonical code is highly error-robust (\citealt{Freeland1998a}). It was also suggested that this property might be explained as a result of non-adaptive evolution (\citealt{Massey2016}). Whatever the case, this feature makes sense in terms of the code functionality, so we regard it, together with block structure, as essentially model-independent (which is not strictly correct, as degree of error-robustness depends on how one defines related amino acid properties). We also note that the set of canonical amino acids reveals properties strongly suggesting that this set had been selected adaptively (\citealt{Philip2011,Ilardo2015}). Although this data leaves the stereochemical and biosynthetic models considered below in certain doubt, by itself it has nothing to say about adaptive (or non-adaptive) origin of error-robustness, because the latter is defined by the mapping of the code, not by the amino acid set alone.

To generate error-robust codes, we first populate blocks with amino acids randomly, then calculate the value $\phi$ of the cost function of the resulting code as described in \citet{Novozhilov2007} using polar requirement scale as a measure, and leave in the sample only those codes which have cost function value not larger than ${\phi}_0 + \sigma$, where $\sigma$ is the standard deviation in the $\phi$-distribution of all codes, and ${\phi}_0$ is the value of the canonical code. Since ${\phi}_0$ is lower than the mode of the distribution by ${\sim}4{\sigma}$, even the value ${\phi}_0 + \sigma$ still represents highly robust codes.
\vskip4pt

{\bf Model 2: Stereochemical hypothesis.} According to the stereochemical model, the assignments of the canonical code were dictated by physicochemical affinities between amino acids and their cognate codons (or anticodons). This model finds support in SELEX experiments (\citealt{Yarus2009}), where RNA aptamers evolved {\it in vitro} to bind some amino acids tend to be enriched with codons or anticodons cognate to those amino acids. However, the major problem with the stereochemical model is that these experimental results come from complex amino acids unlikely to have been available at the time of code emergence (\citealt{Koonin2017}). All in all, whether or not physicochemical affinities were responsible for the assignments in the code is irrelevant for its current functionality, since amino acids interact with codons exclusively via tRNAs, so interpretation of the SELEX data is model-dependent.

To generate codes complying with the stereochemical argument, we populate the blocks in the following manner. The nine amino acids for which RNA-binding data are available (\citealt{Yarus2009}) are assigned, with fixed probabilities, to the blocks which embrace the majority (or all) of codons cognate to those amino acids according to the canonical code, while the rest of amino acids are populated randomly among the rest of the blocks. Specifically, we map the dissociation constants $K_d$ to assignment probabilities as $p = -0.036 \ln(K_d / 1.0$ M$ ) + 0.5$  (where $K_d$ is expressed in molar units), so that, e.g., probability for Phe to be assigned to TTY (or TTN, or TTH, depending on the generated block structure) is 0.86 (see Table~\ref{table:stereodata}). We also checked the ``deterministic'' version of this model, where all probabilities in Table~\ref{table:stereodata} are set to 1, i.e. all nine amino acids are rigidly fixed within cognate codon boxes.
\vskip4pt

{\bf Model 3: Biosynthetic hypothesis.} According to the biosynthetic model, the assignments of the canonical code were guided by evolution of amino acid biosynthesis pathways. This model finds support in the pattern of distribution of precursor-product amino acid pairs observed in the real code (Table~\ref{table:biosynthetic}), where codons of these pairs tend to be separated by only a single point mutation (\citealt{Wong1975,DiGiulio2004}). Interpretation of this observation is also model-dependent (besides, this model has been criticized for the lack of statistical robustness, see \citealt{Amirnovin1997,Ronneberg2000}).

\begin{table}
	\begin{center}
		\caption{The set of precursor-product pairs (according to \citealt{Wong1975}). \\ Italicized amino acids are those which do not represent products to other canonical amino acids in biosynthetic pathways.} 
		\begin{tabular}{cccc}
			\hhline{====}
			{\it Asp} $\to$ Thr &  {\it Glu} $\to$ Gln & {\it Ser} $\to$ Trp & Thr $\to$ Ile \\
			{\it Asp} $\to$ Asn &  {\it Glu} $\to$ Arg & {\it Ser} $\to$ Cys & Thr $\to$ Met \\
			{\it Asp} $\to$ Lys &  {\it Glu} $\to$ Pro & {\it Phe} $\to$ Tyr & Gln $\to$ His \\
			\multicolumn{4}{l}{ {\it Val} $\to$ Leu } \\
			\hline
		\end{tabular}
		\label{table:biosynthetic}
	\end{center}
\end{table}

To generate genetic codes complying with the biosynthetic argument, we populate the blocks in the following manner. First, five ultimate precursors (marked with italic in Table~\ref{table:biosynthetic}) are assigned to five random blocks. Then, for each of them products are introduced sequentially in such a way that they differ from their precursors only in one codon position (in which of the three is determined randomly with equal probabilities). After that Ala and Gly (which are not in simple precursor-product pair relationships), as well as stop-signal, are introduced into the code. Finally, if empty blocks are still left, they are populated with amino acids chosen randomly.

\vskip4pt
{\bf Result of the statistical test.} Some authors note that the three major models of the code evolution are not mutually exclusive (\citealt{Knight1999a,Koonin2009}), with stereochemical affinities being at work at the initial stage, with biosynthetic expansion then taking over, and with selection for error minimization making the final optimization. Therefore, one should test, among others, the combination of models 1+2+3. However, while the models themselves are indeed not mutually exclusive, the actual data employed to support stereochemical and biosynthetic models are hardly compatible. It would be natural to expect that amino acids assigned according to stereochemical argument would serve as precursors in further biosynthetic expansion of the code. However, the opposite is true~-- except Phe and Val, all amino acids from Table~\ref{table:stereodata} represent products in Table~\ref{table:biosynthetic}. Therefore, it is impossible to generate codes complying with models 2 and 3 simultaneously, but we check all other combinations. Besides, we also make the test in which codes have block structure but are not subject to other assumptions. The results of all these tests are shown in Table~\ref{table:result}. It might be seen from the table that the fraction of codes scoring at least $S_0 = 30.5$ does not show significant variation between null hypotheses, and that the symmetries of strings in Fig.~\ref{fig:codogram}d cannot be ascribed to either individual or combined effect of model-dependent and model-independent biological aspects.

We have also probed tentatively other levels in the hierarchy of symmetries. The symmetrized version of the code shares with the canonical code three quantitative symmetries in Rumer's bisection and in codon sortings based on the type of the first/second nucleotide. As mentioned above, these sortings apply to any genetic code and therefore, provided that all combinations of subgroups within them are taken into account, it is statistically correct to ask what fraction of codes reveals at least the same number of quantitative symmetries within those sortings (with that, proline might be considered either as it is or with a nucleon transferred~-- since this transfer is applied universally once and for all, it does not change the degrees of freedom and probabilities). To this end, we had performed Test 0 (since it is the least time-consuming) over the sample of $10^{10}$ codes, and checked what fraction of codes had $S \ge S_0$ and at least three quantitative symmetries within those sortings. We found that while there were seven codes with $S \ge S_0$ and two nucleon balances, there were none with three balances. Thus, $10^{-10}$ represents an upper-bound probabilistic estimate. The computer code (written in C++ using Qt 5) developed for this analysis is available at \url{https://sourceforge.net/projects/gencodegenerator}.

\def\arraystretch{1.4}
\begin{table}[H]
	\begin{center}
		\caption{Fraction of codes with $S \ge S_0$ \\ (the sample size in each case is $N_{total} = 5\times 10^8$ codes).} 
		\begin{tabular}{cc}
			\hhline{==}
			Null hypothesis &  $N_{S \ge S_0} / N_{total}$ \\
			\hline
			0		&	$6.1 \times 10^{-6}$ \\
			0+1		&	$5.6 \times 10^{-6}$ \\
			0+2		&	$7.9 \times 10^{-6}$ \\
			0+3		&	$6.4 \times 10^{-6}$ \\
			0+1+2	&	$8.2 \times 10^{-6}$ \\
			0+1+3	&	$5.8 \times 10^{-6}$ \\
			0+2*	&	$8.8 \times 10^{-6}$ \\
			0+1+2*	&	$8.5 \times 10^{-6}$ \\
			\hline 
			\label{table:result}
		\end{tabular} 
		
		Null hypotheses: 0~-- block structure; 1~-- error minimization model; 2~-- stereochemical model (probabilistic); 2*~-- stereochemical model (deterministic); 3~-- biosynthetic model.
	\end{center}
\end{table}

\bigskip
\bigskip
\hypertarget{videos}{\,}
\begin{tcolorbox}[notitle,no shadow,before upper={\parindent10pt},float,floatplacement=b]
\section*{\small Video presentations}
\bigskip
\noindent \textbf{Video 1:} Description of the attributes encoded with the \mbox{strategy} ``from amino acids to codons'' [05 min 27 sec]
\medskip

\noindent \textbf{Video 2:} Description of the attributes encoded with the \mbox{strategy} ``from codons to amino acids'' [11 min 03 sec]

\bigskip
\bigskip

\noindent The videos might be accesses at:
\bigskip

\textbf{YouTube:}
\smallskip

\qquad Video 1: \url{https://youtu.be/VxnyjxtLfsA}

\qquad Video 2: \url{https://youtu.be/3bxZZlTfSsY}

\bigskip
\medskip

\textbf{bioSETI website}:
\smallskip

\qquad \url{https://bioseti.info/videos/}

\bigskip
\medskip

\textbf{Publisher's website}:

\qquad \url{https://doi.org/10.1017/S1473550417000210}
\end{tcolorbox}


\newpage

\begin{thebibliography}{116}
	\expandafter\ifx\csname natexlab\endcsname\relax\def\natexlab#1{#1}\fi
	\expandafter\ifx\csname url\endcsname\relax
	\def\url#1{{\tt #1}}\fi
	\expandafter\ifx\csname urlprefix\endcsname\relax\def\urlprefix{URL }\fi
	
	
	\fontsize{8.4}{8.6}
	\selectfont
	
	\bibitem[{Ahituv et~al.(2007)Ahituv, Zhu, Visel, Holt, Afzal, Pennacchio \&
		Rubin}]{Ahituv2007}
	Ahituv, N., Zhu, Y., Visel, A., Holt, A., Afzal, V., Pennacchio, L.~A. \&
	Rubin, E.~M. (2007).
	\newblock {\href{http://dx.doi.org/10.1371/journal.pbio.0050234}{Deletion of ultraconserved elements yields viable mice.}}
	\newblock {\em PLoS Biol.\/}, {\em 5\/}, e234.
	
	\bibitem[{Amirnovin(1997)}]{Amirnovin1997}
	Amirnovin, R. (1997).
	\newblock {\href{http://dx.doi.org/10.1007/PL00006170}{An analysis of the metabolic theory of the origin of the genetic
			code.}}
	\newblock {\em J. Mol. Evol.\/}, {\em 44\/}, 473--476.
	
	\bibitem[{Anglada-Escude et~al.(2014)Anglada-Escude, Arriagada, Tuomi,
		Zechmeister, Jenkins, Ofir, Dreizler, Gerlach, Marvin, Reiners, Jeffers,
		Butler, Vogt, Amado, Rodriguez-Lopez, Berdinas, Morin, Crane, Shectman,
		Thompson, Diaz, Rivera, Sarmiento \& Jones}]{Anglada-Escude2014}
	Anglada-Escude, G., Arriagada, P., Tuomi, M., Zechmeister, M., Jenkins, J.~S.,
	Ofir, A., Dreizler, S., Gerlach, E., Marvin, C.~J., Reiners, A., Jeffers,
	S.~V., Butler, R.~P., Vogt, S.~S., Amado, P.~J., Rodriguez-Lopez, C.,
	Berdinas, Z.~M., Morin, J., Crane, J.~D., Shectman, S.~A., Thompson, I.~B.,
	Diaz, M., Rivera, E., Sarmiento, L.~F. \& Jones, H. R.~A. (2014).
	\newblock {\href{http://dx.doi.org/10.1093/mnrasl/slu076}{Two planets around Kapteyn's star: a cold and a temperate
			super-Earth orbiting the nearest halo red dwarf.}}
	\newblock {\em Mon. Not. R. Astron. Soc. Let.\/},
	{\em 443\/}, L89--L93.
	
	\bibitem[{Bains \&  Schulze-Makuch(2016)}]{Bains2016}
	Bains, W. \&  Schulze-Makuch, D. (2016).
	\newblock {\href{http://dx.doi.org/10.3390/life6030025}{The cosmic zoo: the (near) inevitability of the
			evolution of complex, macroscopic life.}}
	\newblock {\em Life\/}, {\em 6\/}, 25.  
	
	\bibitem[{Baranov et~al.(2015)Baranov, Atkins \& Yordanova}]{Baranov2015}
	Baranov, P.~V., Atkins, J.~F. \& Yordanova, M.~M. (2015).
	\newblock {\href{http://dx.doi.org/10.1038/nrg3963}{Augmented genetic decoding: global, local and temporal alterations
			of decoding processes and codon meaning.}}
	\newblock {\em Nat. Rev. Genet.\/}, {\em 16\/}, 517--529.
	
	\bibitem[{Bell et~al.(2015)Bell, Boehnke, Harrison \& Mao}]{Bell2015}
	Bell, E.~A., Boehnke, P., Harrison, T.~M. \& Mao, W.~L. (2015).
	\newblock {\href{http://dx.doi.org/10.1073/pnas.1517557112}{Potentially biogenic carbon preserved in a 4.1 billion-year-old
			zircon.}}
	\newblock {\em Proc. Natl. Acad. Sci. USA\/}, {\em 112\/},
	14518--14521.
	
	\bibitem[{Brar(2016)}]{Brar2016}
	Brar, G.~A. (2016).
	\newblock {\href{http://dx.doi.org/10.1016/j.cell.2016.09.022}{Beyond the triplet code: context cues transform translation.}}
	\newblock {\em Cell\/}, {\em 167\/},
	1681--1692.
	
	\bibitem[{Campante et~al.(2015)Campante, Barclay, Swift, Huber, Adibekyan,
		Cochran, Burke, Isaacson, Quintana, Davies, {Silva Aguirre}, Ragozzine,
		Riddle, Baranec, Basu, Chaplin, Christensen-Dalsgaard, Metcalfe, Bedding,
		Handberg, Stello, Brewer, Hekker, Karoff, Kolbl, Law, Lundkvist, Miglio,
		Rowe, Santos, {Van Laerhoven}, Arentoft, Elsworth, Fischer, Kawaler,
		Kjeldsen, Lund, Marcy, Sousa, Sozzetti \& White}]{Campante2015}
	Campante, T.~L., Barclay, T., Swift, J.~J., Huber, D., Adibekyan, V.~Z.,
	Cochran, W., Burke, C.~J., Isaacson, H., Quintana, E.~V., Davies, G.~R.,
	{Silva Aguirre}, V., Ragozzine, D., Riddle, R., Baranec, C., Basu, S.,
	Chaplin, W.~J., Christensen-Dalsgaard, J., Metcalfe, T.~S., Bedding, T.~R.,
	Handberg, R., Stello, D., Brewer, J.~M., Hekker, S., Karoff, C., Kolbl, R.,
	Law, N.~M., Lundkvist, M., Miglio, A., Rowe, J.~F., Santos, N.~C., {Van
		Laerhoven}, C., Arentoft, T., Elsworth, Y.~P., Fischer, D.~A., Kawaler,
	S.~D., Kjeldsen, H., Lund, M.~N., Marcy, G.~W., Sousa, S.~G., Sozzetti, A.,
	\& White, T.~R. (2015).
	\newblock {\href{http://dx.doi.org/10.1088/0004-637X/799/2/170}{An ancient extrasolar system with five sub-Earth-size planets.}}
	\newblock {\em Astrophys. J.\/}, {\em 799\/}, 170.
	
	\bibitem[{Chin(2014)}]{Chin2014}
	Chin, J.~W. (2014).
	\newblock {\href{http://dx.doi.org/10.1146/annurev-biochem-060713-035737}{Expanding and reprogramming the genetic code of cells and animals.}}
	\newblock {\em Annu. Rev. Biochem.\/}, {\em 83\/}, 379--408.
	
	\bibitem[{Chrisomalis(2010)}]{Chrisomalis2010}
	Chrisomalis, S. (2010).
	\newblock {\em {\href{https://books.google.com/books?id=kXZhBAAAQBAJ}{Numerical Notation: A Comparative History.}}\/}
	\newblock Cambridge: Cambridge Univ. Press.
	
	\bibitem[{{\'{C}}irkovi{\'{c}}(2014)}]{Cirkovic2014}
	{\'{C}}irkovi{\'{c}}, M.~M. (2014).
	\newblock {\href{http://dx.doi.org/10.1007/s10539-013-9397-8}{Evolutionary contingency and SETI revisited.}}
	\newblock {\em Biol. Philos.\/}, {\em 29\/}, 539--557.
	
	\bibitem[{{Conway Morris}(2003{\natexlab{a}})}]{ConwayMorris2003a}
	{Conway Morris}, S. (2003{\natexlab{a}}).
	\newblock {\em {\href{https://books.google.com/books?id=EdQLAQAAQBAJ}{Life's Solution: Inevitable Humans in a Lonely Universe.}}\/}
	\newblock Cambridge: Cambridge Univ. Press.
	
	\bibitem[{{Conway Morris}(2003{\natexlab{b}})}]{ConwayMorris2003b}
	{Conway Morris}, S. (2003{\natexlab{b}}).
	\newblock {\href{http://dx.doi.org/10.1017/S1473550403001460}{The navigation of biological hyperspace.}}
	\newblock {\em Int. J. Astrobiol.\/}, {\em 2\/}, 149--152.
	
	\bibitem[{{Conway Morris}(2011)}]{ConwayMorris2011}
	{Conway Morris}, S. (2011).
	\newblock {\href{http://dx.doi.org/10.1098/rsta.2010.0276}{Predicting what extra-terrestrials will be like: and preparing for
			the worst.}}
	\newblock {\em Phil. Trans. R. Soc. A\/}, {\em
		369\/}, 555--571.
	
	\bibitem[{Crick(1981)}]{Crick1981}
	Crick, F. H.~C. (1981).
	\newblock {\em {Life Itself: Its Origin and Nature.}\/}
	\newblock New York: Simon {\&} Schuster.
	
	\bibitem[{Crick \& Orgel(1973)}]{Crick1973}
	Crick, F. H.~C. \& Orgel, L.~E. (1973).
	\newblock {\href{http://dx.doi.org/10.1016/0019-1035(73)90110-3}{Directed panspermia.}}
	\newblock {\em Icarus\/}, {\em 19\/}, 341--346.
	
	\bibitem[{Danckwerts \& Neubert(1975)}]{Danckwerts1975}
	Danckwerts, H.~J. \& Neubert, D. (1975).
	\newblock {\href{http://dx.doi.org/10.1007/BF01732219}{Symmetries of genetic code-doublets.}}
	\newblock {\em J. Mol. Evol.\/}, {\em 5\/}, 327--332.
	
	\bibitem[{Dantzig(1954)}]{Dantzig1954}
	Dantzig, T. (1954).
	\newblock {\em {\href{https://books.google.kz/books?id=Pg_RKtlVlNMC}{Number: The Language of Science.}}\/}
	\newblock New York: Simon {\&} Schuster, 4th ed.
	
	\bibitem[{Davies(2012)}]{Davies2011a}
	Davies, P. C.~W. (2012).
	\newblock {\href{http://dx.doi.org/10.1016/j.actaastro.2011.06.022}{Footprints of alien technology.}}
	\newblock {\em Acta Astronaut.\/}, {\em 73\/}, 250--257.
	
	\bibitem[{Davis(1996)}]{Davis1996}
	Davis, J. (1996).
	\newblock {\href{http://dx.doi.org/10.1080/00043249.1996.10791743}{Microvenus.}}
	\newblock {\em Art J.\/}, {\em 55\/}, 70--74.
	
	\bibitem[{Dawkins(2004)}]{Dawkins2004}
	Dawkins, R. (2004).
	\newblock {\em {\href{https://books.google.com/books?id=Scxg1Pu3dOsC}{The Ancestor's Tale: A Pilgrimage to the Dawn of Life.}}\/}
	\newblock Boston: Houghton Mifflin Harcourt.
	
	\bibitem[{{de Grijs}(2010)}]{deGrijs2010}
	de Grijs, R. (2010).
	\newblock {\href{http://dx.doi.org/10.1098/rsta.2009.0253}{A revolution in star cluster research: setting the scene.}}
	\newblock {\em Phil. Trans. R. Soc. A\/}, {\em 368\/}, 693-711.
	
	\bibitem[{Deacon(1997)}]{Deacon1997}
	Deacon, T.~W. (1997).
	\newblock {\em {\href{https://books.google.com/books?id=u6IXngEACAAJ}{The Symbolic Species: The Co-evolution of Language and the
				Brain.}}\/}
	\newblock New York: W.W. Norton \& Co.
	
	\bibitem[{{Di Giulio}(2004)}]{DiGiulio2004}
	{Di Giulio}, M. (2004).
	\newblock {\href{http://dx.doi.org/10.1016/j.plrev.2004.05.001}{The coevolution theory of the origin of the genetic code.}}
	\newblock {\em Phys. Life Rev.\/}, {\em 1\/}, 128--137.
	
	\bibitem[{Doolittle(2000)}]{Doolittle2000}
	Doolittle, W.~F. (2000).
	\newblock {\href{http://dx.doi.org/10.1016/S0959-440X(00)00096-8}{The nature of the universal ancestor and the evolution of the proteome.}}
	\newblock {\em  Curr. Opin. Struct. Biol.\/}, {\em 10\/}, 355--358.
	
	\bibitem[{Forterre \& Philippe(1999)}]{Forterre1999}
	Forterre, P. \& Philippe, H. (1999).
	\newblock {\href{http://dx.doi.org/10.1002/(SICI)1521-1878(199910)21:10<871::AID-BIES10>3.0.CO;2-Q}{Where is the root of the universal tree of life?}}
	\newblock {\em BioEssays\/}, {\em 21\/}, 871--879.
	
	\bibitem[{Fournier \& Gogarten(2007)}]{Fournier2007}
	Fournier, G.~P. \& Gogarten, J.~P. (2007).
	\newblock {\href{http://dx.doi.org/10.1007/s00239-007-9024-x}{Signature of a primitive genetic code in ancient protein lineages.}}
	\newblock {\em J. Mol. Evol.\/}, {\em 65\/}, 425--436.
	
	\bibitem[{Freeland \& Hurst(1998)}]{Freeland1998a}
	Freeland, S.~J. \& Hurst, L.~D. (1998).
	\newblock {\href{http://dx.doi.org/10.1007/PL00006381}{The genetic code is one in a million.}}
	\newblock {\em J. Mol. Evol.\/}, {\em 47\/}, 238--248.
	
	\bibitem[{Freitas(1983)}]{Freitas1983}
	Freitas, R.~A. (1983).
	\newblock {\href{http://adsabs.harvard.edu/abs/1983JBIS...36..501F}{The search for extraterrestrial artifacts (SETA).}}
	\newblock {\em J. Brit. Interplanet. Soc.\/}, {\em 36\/},
	501--506.
	
	\bibitem[{Gazal{\'{e}}(2000)}]{Gazale2000}
	Gazal{\'{e}}, M. (2000).
	\newblock {\em {\href{https://books.google.com/books?id=WO1gQYIrG24C}{Number: From Ahmes to Cantor.}}\/}
	\newblock Princeton: Princeton Univ. Press.
	
	\bibitem[{Gibbs(2001)}]{Gibbs2001}
	Gibbs, W.~W. (2001).
	\newblock {\href{http://dx.doi.org/10.1038/scientificamerican0401-40}{Art as a form of life.}}
	\newblock {\em Sci. Am.\/}, {\em 284\/}, 37--39.
	
	\bibitem[{Gibson et~al.(2010)}]{Gibson2010}
	Gibson, D.~G., Glass, J.~I., Lartigue, C., Noskov, V.~N., Chuang, R.-Y., Algire, M.~A., Benders, G.~A., Montague, M.~G., Ma, L., Moodie, M.~M., Merryman, C., Vashee, S., Krishnakumar, R., Assad-Garcia, N., Andrews-Pfannkoch, C., Denisova, E.~A., Young, L., Qi, Z.-Q., Segall-Shapiro, T.~H., Calvey, C.~H., Parmar, P.~P., Hutchison, C.~A., Smith, H.~O., Venter, J.~C. (2010).
	\newblock {\href{http://dx.doi.org/10.1126/science.1190719}{Creation of a bacterial cell controlled by a chemically synthesized genome.}}
	\newblock {\em Science\/}, {\em 329\/}, 52--56.
	
	\bibitem[{Gieg{\'{e}}(2008)}]{Giege2008}
	Gieg{\'{e}}, R. (2008).
	\newblock {\href{http://dx.doi.org/10.1038/nsmb.1498}{Toward a more complete view of tRNA biology.}}
	\newblock {\em Nat. Struct. Mol. Biol.\/}, {\em 15\/}, 1007--1014.
	
	\bibitem[{Glansdorff et~al.(2008)Glansdorff, Xu \& Labedan}]{Glansdorff2008}
	Glansdorff, N., Xu, Y. \& Labedan, B. (2008).
	\newblock {\href{http://dx.doi.org/10.1186/1745-6150-3-29}{The last universal common ancestor: emergence, constitution and
			genetic legacy of an elusive forerunner.}}
	\newblock {\em Biol. Dir.\/}, {\em 3\/}, 29.
	
	\bibitem[{Gould(1985)}]{Gould1985}
	Gould, S.~J. (1985).
	\newblock {\em {The Flamingo's Smile: Reflections in Natural History.}\/}
	\newblock New York: Norton {\&} Co.
	
	\bibitem[{Gould(1989)}]{Gould1989}
	Gould, S.~J. (1989).
	\newblock {\em {Wonderful life: The Burgess Shale and the Nature of
			History.}\/}
	\newblock New York: Norton {\&} Co.
	
	\bibitem[{Gros(2016)}]{Gros2016}
	Gros, C. (2016).
	\newblock {\href{http://dx.doi.org/10.1007/s10509-016-2911-0}{Developing ecospheres on transiently habitable planets: the genesis project.}\/}
	\newblock {\em Astrophys. Space Sci.\/}, {\em 361\/}, 324.
	
	\bibitem[{Guo \& Schimmel(2013)}]{Guo2013}
	Guo, M. \& Schimmel, P. (2013).
	\newblock {\href{http://dx.doi.org/10.1038/nchembio.1158}{Essential nontranslational functions of tRNA synthetases.}}
	\newblock {\em Nat. Chem. Biol.\/}, {\em 9\/}, 145--153.
	
	\bibitem[{Haldane(1954)}]{Haldane1954}
	Haldane, J. B.~S. (1954).
	\newblock {The origins of life.}
	\newblock {\em New Biol.\/}, {\em 16\/}, 12--27.
	
	\bibitem[{Hasegawa \& Miyata(1980)}]{Hasegawa1980}
	Hasegawa, M. \& Miyata, T. (1980).
	\newblock {\href{http://dx.doi.org/10.1007/BF00928404}{On the antisymmetry of the amino acid code table.}}
	\newblock {\em Orig. Life\/}, {\em 10\/}, 265--270.
	
	\bibitem[{Hayes(1998)}]{Hayes1998}
	Hayes, B. (1998).
	\newblock {\href{https://www.americanscientist.org/issues/num2/1998/1/the-invention-of-the-genetic-code/1}{The invention of the genetic code.}}
	\newblock {\em Am. Sci.\/}, {\em 86\/}, 8--14.
	
	\bibitem[{Hoch \& Losick(1997)}]{Hoch1997}
	Hoch, J.~A. \& Losick, R. (1997).
	\newblock {\href{http://dx.doi.org/10.1038/36747}{Panspermia, spores and the Bacillus subtilis genome.}}
	\newblock {\em Nature\/}, {\em 390\/}, 237--238.
	
	\bibitem[{Hoffman et~al.(1995)Hoffman, Anderson, DuBois \&
		Prescott}]{Hoffman1995}
	Hoffman, D.~C., Anderson, R.~C., DuBois, M.~L. \& Prescott, D.~M. (1995).
	\newblock {\href{http://dx.doi.org/10.1093/nar/23.8.1279}{Macronuclear gene-sized molecules of hypotrichs.}}
	\newblock {\em Nucleic Acids Res.\/}, {\em 23\/}, 1279--1283.
	
	\bibitem[{Ifrah(2000)}]{Ifrah2000}
	Ifrah, G. (2000).
	\newblock {\em {\href{https://books.google.com/books?id=wxClQgAACAAJ}{The Universal History of Numbers: From Prehistory to the
				Invention of the Computer.}}\/}
	\newblock New York: John Wiley {\&} Sons.
	
	\bibitem[{Ilardo et~al.(2015)Ilardo, Meringer, Freeland, Rasulev \& {Cleaves
			II}}]{Ilardo2015}
	Ilardo, M., Meringer, M., Freeland, S.~J., Rasulev, B. \& {Cleaves II}, H.~J.
	(2015).
	\newblock {\href{http://dx.doi.org/10.1038/srep09414}{Extraordinarily adaptive properties of the genetically encoded amino
			acids.}}
	\newblock {\em Sci. Rep.\/}, {\em 5\/}, 9414.
	
	\bibitem[{Isenbarger et~al.(2008)Isenbarger, Carr, Johnson, Finney, Church,
		Gilbert, Zuber \& Ruvkun}]{Isenbarger2008}
	Isenbarger, T.~A., Carr, C.~E., Johnson, S.~S., Finney, M., Church, G.~M.,
	Gilbert, W., Zuber, M.~T. \& Ruvkun, G. (2008).
	\newblock {\href{http://dx.doi.org/10.1007/s11084-008-9148-z}{The most conserved genome segments for life detection on Earth and
			other planets.}}
	\newblock {\em Orig. Life Evol. Biosph.\/}, {\em 38\/},
	517--533.
	
	\bibitem[{Ivanova et~al.(2014)Ivanova, Schwientek, Tripp, Rinke, Pati,
		Huntemann, Visel, Woyke, Kyrpides \& Rubin}]{Ivanova2014}
	Ivanova, N.~N., Schwientek, P., Tripp, H.~J., Rinke, C., Pati, A., Huntemann,
	M., Visel, A., Woyke, T., Kyrpides, N.~C. \& Rubin, E.~M. (2014).
	\newblock {\href{http://dx.doi.org/10.1126/science.1250691}{Stop codon reassignments in the wild.}}
	\newblock {\em Science\/}, {\em 344\/}, 909--913.
	
	\bibitem[{Kaplan(1999)}]{Kaplan1999}
	Kaplan, R. (1999).
	\newblock {\em {\href{https://books.google.com/books?id=ZtDnCwAAQBAJ}{The Nothing That Is: A Natural History of Zero.}}\/}
	\newblock Oxford: Oxford Univ. Press.
	
	\bibitem[{Knight et~al.(1999)Knight, Freeland \& Landweber}]{Knight1999a}
	Knight, R.~D., Freeland, S.~J. \& Landweber, L.~F. (1999).
	\newblock {\href{http://dx.doi.org/10.1016/S0968-0004(99)01392-4}{Selection, history and chemistry: the three faces of the genetic
			code.}}
	\newblock {\em Trends Biochem. Sci.\/}, {\em 24\/}, 241--247.
	
	\bibitem[{Knight et~al.(2001)Knight, Freeland \& Landweber}]{Knight2001a}
	Knight, R.~D., Freeland, S.~J. \& Landweber, L.~F. (2001).
	\newblock {\href{http://dx.doi.org/10.1038/35047500}{Rewiring the keyboard: evolvability of the genetic code.}}
	\newblock {\em Nat. Rev. Genet.\/}, {\em 2\/}, 49--58.
	
	\bibitem[{Koonin(2011)}]{Koonin2011b}
	Koonin, E.~V. (2011).
	\newblock {\em {\href{https://books.google.com/books?id=fvmv2kU6PrYC}{The Logic of Chance: The Nature and Origin of Biological
				Evolution.}}\/}
	\newblock Upper Saddle River: FT Press.

	\bibitem[{Koonin \& Novozhilov(2009)}]{Koonin2009}
	Koonin, E.~V. \& Novozhilov, A.~S. (2009).
	\newblock {\href{http://dx.doi.org/10.1002/iub.146}{Origin and evolution of the genetic code: the universal enigma.}}
	\newblock {\em IUBMB Life\/}, {\em 61\/}, 99--111.

	\bibitem[{Koonin \& Novozhilov(2017)}]{Koonin2017}
	Koonin, E.~V. \& Novozhilov, A.~S. (2017).
	\newblock {\href{http://dx.doi.org/10.1146/annurev-genet-120116-024713}{Origin and evolution of the universal genetic code.}}
	\newblock {\em Annu. Rev. Genet.\/}, {\em 51\/}.
	
	\bibitem[{Kouwenhoven et~al.(2016)}]{Kouwenhoven2016}
	Kouwenhoven, M.~B.~N., Shu, Q., Cai, M.~X. \& Spurzem, R. (2016).
	\newblock {\href{https://arxiv.org/abs/1609.00898}{Planetary systems in star clusters.}}
	\newblock {\em Proceedings of Cosmic-Lab: Star Clusters as Cosmic Laboratories for Astrophysics, Dynamics, and Fundamental Physics - MODEST16\/}.
	
	\bibitem[{Lagerkvist(1978)}]{Lagerkvist1978}
	Lagerkvist, U. (1978).
	\newblock {\href{http://dx.doi.org/10.1073/pnas.75.4.1759}{``Two out of three'': an alternative method for codon reading.}}
	\newblock {\em Proc. Natl. Acad. Sci. USA\/},
	{\em 75\/}, 1759--1762.
	
	\bibitem[{Lajoie et~al.(2013)Lajoie, Kosuri, Mosberg, Gregg, Zhang \&
		Church}]{Lajoie2013a}
	Lajoie, M.~J., Kosuri, S., Mosberg, J.~A., Gregg, C.~J., Zhang, D. \& Church,
	G.~M. (2013).
	\newblock {\href{http://dx.doi.org/10.1126/science.1241460}{Probing the limits of genetic recoding in essential genes.}}
	\newblock {\em Science\/}, {\em 342\/}, 361--363.
	
	\bibitem[{Lajoie et~al.(2016)Lajoie, S{\"{o}}ll \& Church}]{Lajoie2016}
	Lajoie, M.~J., S{\"{o}}ll, D. \& Church, G.~M. (2016).
	\newblock {\href{http://dx.doi.org/10.1016/j.jmb.2015.09.003}{Overcoming challenges in engineering the genetic code.}}
	\newblock {\em J. Mol. Biol.\/}, {\em 428\/}, 1004--1021.
	
	\bibitem[{Lemarchand \& Lomberg(2009)}]{Lemarchand2009}
	Lemarchand, G. \& Lomberg, J. (2009).
	\newblock {\href{http://dx.doi.org/10.1162/leon.2009.42.5.396}{Universal cognitive maps and the search for intelligent life in the
			universe.}}
	\newblock {\em Leonardo\/}, {\em 42\/}, 396--402.
	
	\bibitem[{Lineweaver(2001)}]{Lineweaver2001}
	Lineweaver, C.H. (2001).
	\newblock {\href{http://dx.doi.org/10.1006/icar.2001.6607}{An estimate of the age distribution of terrestrial planets in the universe:
			quantifying metallicity as a selection effect.}}
	\newblock {\em Icarus\/}, {\em 151\/}, 307--313.
	
	\bibitem[{Liu \& Schultz(2010)}]{Liu2010}
	Liu, C.~C. \& Schultz, P.~G. (2010).
	\newblock {\href{http://dx.doi.org/10.1146/annurev.biochem.052308.105824}{Adding new chemistries to the genetic code.}}
	\newblock {\em Annu. Rev. Biochem.\/}, {\em 79\/}, 413--444.
	
	\bibitem[{Lobkovsky \& Koonin(2012)}]{Lobkovsky2012}
	Lobkovsky, A.~E. \& Koonin, E.~V. (2012).
	\newblock {\href{http://dx.doi.org/10.3389/fgene.2012.00246}{Replaying the tape of life: quantification of the predictability of
			evolution.}}
	\newblock {\em Front. Genet.\/}, {\em 3\/}, 1--8.
	
	\bibitem[{Lotman et~al.(1978)}]{Lotman1978}
	Lotman, Yu.~M., Uspensky, B.~A. \& Mihaychuk, G. (1978).
	\newblock {\href{http://dx.doi.org/10.2307/468571}{On the semiotic mechanism of culture.}}
	\newblock {\em New Lit. Hist.\/}, {\em 9\/}, 211--232.
	
	\bibitem[{Makukov \& shCherbak(2014)}]{Makukov2014}
	Makukov, M.~A. \& shCherbak, V.~I. (2014).
	\newblock {\href{http://dx.doi.org/10.1016/j.lssr.2014.07.003}{Space ethics to test directed panspermia.}}
	\newblock {\em Life Sci. Space Res.\/}, {\em 3\/}, 10--17.
	
	\bibitem[{Maraia \& Iben(2014)}]{Maraia2014}
	Maraia, R.~J. \& Iben, J.~R. (2014).
	\newblock {\href{http://dx.doi.org/10.1261/rna.044115.113.2}{Different types of secondary information in the genetic code.}}
	\newblock {\em RNA\/}, {\em 20\/}, 977--984.
	
	\bibitem[{Mart{\'{i}}nez-Barbosa et~al.(2016)}]{MartinezBarbosa2016}
	Mart{\'{i}}nez-Barbosa, C.~A., Brown, A.~G.~A., Boekholt, T., {Portegies Zwart}, S., Antiche, E., Antoja, T. (2016).
	\newblock {\href{https://dx.doi.org/10.1093/mnras/stw006}{The evolution of the Sun's birth cluster and the search for the solar siblings with Gaia.}}
	\newblock {\em Mon. Not. R. Astron. Soc.\/}, {\em 457\/}, 1062--1075.
	
	\bibitem[{Marx(1979)}]{Marx1979}
	Marx, G. (1979).
	\newblock {\href{http://dx.doi.org/10.1016/0094-5765(79)90158-9}{Message through time.}}
	\newblock {\em Acta Astronaut.\/}, {\em 6\/}, 221--225.
	
	\bibitem[{Marx(1986)}]{Marx1986}
	Marx, G. (1986).
	\newblock {The problem of simultaneity}.
	\newblock In V.~A. Ambartsumian, N.~S. Kardashev \& V.~S. Troitsky (Eds.) {\em
		The Problem of the Search for Life in the Universe: Proceedings of the SETI
		Symposium in Tallinn\/}, pp. 74--81. Moscow: Nauka.
	\newblock (in Russian).
	
	\bibitem[{Massey(2016)}]{Massey2016}
	Massey, S.~E. (2016).
	\newblock {\href{http://dx.doi.org/10.1016/j.jtbi.2016.08.022}{The neutral emergence of error minimized genetic codes superior to
			the standard genetic code.}}
	\newblock {\em J. Theor. Biol.\/}, {\em 408\/}, 237--242.
	
	\bibitem[{Mautner(1997)}]{Mautner1997}
	Mautner, M.~N. (1997).
	\newblock {\href{http://adsabs.harvard.edu/abs/1997JBIS...50...93M}{Directed panspermia. 3. Strategies and motivation for seeding star-forming clouds.}}
	\newblock {\em J. Brit. Interplanet. Soc.\/}, {\em 50\/}, 93-102.
	
	\bibitem[{Mautner(2000)}]{Mautner2004}
	Mautner, M.~N. (2000).
	\newblock {\em {\href{https://books.google.com/books?id=ctah5ICTy5gC}{Seeding the Universe with Life: Securing Our Cosmological
				Future.}}\/}
	\newblock Weston: Legacy Books.
	
	\bibitem[{Mautner(2009)}]{Mautner2009}
	Mautner, M.~N. (2009).
	\newblock {\href{http://dx.doi.org/10.1111/j.1467-8519.2008.00688.x}{Life-centered ethics, and the human future in space.}}
	\newblock {\em Bioethics\/}, {\em 23\/}, 433-440.
	
	\bibitem[{Mazur(2014)}]{Mazur2014}
	Mazur, J. (2014).
	\newblock {\em {\href{https://books.google.com/books?id=cI4pAgAAQBAJ}{Enlightening Symbols: A Short History of Mathematical Notation
				and its Hidden Powers.}}\/}
	\newblock Princeton: Princeton Univ. Press.
	
	\bibitem[{McGhee(2011)}]{McGhee2011}
	McGhee, G. (2011).
	\newblock {\em {\href{https://books.google.com/books?id=QwDSr1qdqXUC}{Convergent Evolution: Limited Forms Most Beautiful.}}\/}
	\newblock London: The MIT Press.
	
	\bibitem[{Melosh(2003)}]{Melosh2003}
	Melosh, H.~J. (2003).
	\newblock {\href{http://dx.doi.org/10.1089/153110703321632525}{Exchange of meteorites (and life?) between stellar systems.}}
	\newblock {\em Astrobiology\/}, {\em 3\/}, 207--215.
	
	\bibitem[{Meot-Ner \& Matloff(1979)}]{Meot-Ner1979}
	Meot-Ner, M. \& Matloff, G.~L. (1979).
	\newblock {\href{http://adsabs.harvard.edu/abs/1979jbis...32..419m}{Directed panspermia - A technical and ethical evaluation of seeding
			nearby solar systems.}}
	\newblock {\em J. Brit. Interplanet. Soc.\/}, {\em 32\/},
	419--423.
	
	\bibitem[{Meyer et~al.(1991)Meyer, Schmidt, Pl{\"{u}}mper, Hasilik, Mersmann,
		Meyer, Engstr{\"{o}}m \& Heckmann}]{Meyer1991}
	Meyer, F., Schmidt, H.~J., Pl{\"{u}}mper, E., Hasilik, A., Mersmann, G., Meyer,
	H.~E., Engstr{\"{o}}m, A. \& Heckmann, K. (1991).
	\newblock {\href{http://dx.doi.org/10.1073/pnas.88.9.3758}{UGA is translated as cysteine in pheromone 3 of Euplotes
			octocarinatus.}}
	\newblock {\em Proc. Natl. Acad. Sci. USA\/},
	{\em 88\/}, 3758--3761.
	
	\bibitem[{Mileikowsky et~al.(2000)}]{Mileikowsky2000}
	Mileikowsky, C., Cucinotta, F.A., Wilson, J.W., Gladman, B., Horneck, G., Lindegren, L., Melosh, J., Rickman, H., Valtonen, M., Zheng, J.Q. (2000).
	\newblock {\href{http://dx.doi.org/10.1006/icar.1999.6317}{Natural transfer of viable microbes in space 1. From Mars to Earth and Earth to Mars.}}
	\newblock {\em Icarus\/}, {\em 145\/}, 391--427.
	
	\bibitem[{{NC-IUB}(1985)}]{NC-IUB1985}
	{NC-IUB -- Nomenclature Committee of the International Union of Biochemistry}
	(1985).
	\newblock {\href{http://dx.doi.org/10.1093/nar/13.9.3021}{Nomenclature for incompletely specified bases in nucleic acid
			sequences: recommendations 1984.}}
	\newblock {\em Nucleic Acids Res.\/}, {\em 13\/}, 3021--3030.
	
	\bibitem[{Novozhilov et~al.(2007)Novozhilov, Wolf \& Koonin}]{Novozhilov2007}
	Novozhilov, A.~S., Wolf, Y.~I. \& Koonin, E.~V. (2007).
	\newblock {\href{http://dx.doi.org/10.1186/1745-6150-2-24}{Evolution of the genetic code: partial optimization of a random code
			for robustness to translation error in a rugged fitness landscape.}}
	\newblock {\em Biol. Dir.\/}, {\em 2\/}, 24.
	
	\bibitem[{Ostrov et~al.(2016)Ostrov, Landon, Guell, Kuznetsov, Teramoto,
		Cervantes, Zhou, Singh, Napolitano, Moosburner, Shrock, Pruitt, Conway,
		Goodman, Gardner, Tyree, Gonzales, Wanner, Norville, Lajoie \&
		Church}]{Ostrov2016}
	Ostrov, N., Landon, M., Guell, M., Kuznetsov, G., Teramoto, J., Cervantes, N.,
	Zhou, M., Singh, K., Napolitano, M.~G., Moosburner, M., Shrock, E., Pruitt,
	B.~W., Conway, N., Goodman, D.~B., Gardner, C.~L., Tyree, G., Gonzales, A.,
	Wanner, B.~L., Norville, J.~E., Lajoie, M.~J. \& Church, G.~M. (2016).
	\newblock {\href{http://dx.doi.org/10.1126/science.aaf3639}{Design, synthesis, and testing toward a 57-codon genome.}}
	\newblock {\em Science\/}, {\em 353\/}, 819--822.
	
	\bibitem[{Pfalzner(2013)}]{Pfalzner2013}
	Pfalzner, S. (2013).
	\newblock {\href{http://dx.doi.org/10.1051/0004-6361/201218792}{Early evolution of the birth cluster of the solar system.}}
	\newblock {\em Astron. Astrophys.\/}, {\em 549\/}, A82.
	
	\bibitem[{Philip \& Freeland(2011)}]{Philip2011}
	Philip, G.~K. \& Freeland, S.~J. (2011).
	\newblock {\href{http://dx.doi.org/10.1089/ast.2010.0567}{Did evolution select a nonrandom ``alphabet'' of amino acids?}}
	\newblock {\em Astrobiology\/}, {\em 11\/}, 235--240.
	
	\bibitem[{Plotkin \& Kudla(2011)}]{Plotkin2011}
	Plotkin, J.~B. \& Kudla, G. (2011).
	\newblock {\href{http://dx.doi.org/10.1038/nrg2899}{Synonymous but not the same: the causes and consequences of codon bias.}}
	\newblock {\em Nat. Rev. Genet.\/}, {\em 12\/}, 32--42.
	
	\bibitem[{Roberts \& Tesman(2009)}]{Roberts2009}
	Roberts, F.~S. \& Tesman, B. (2009).
	\newblock {\em {\href{https://books.google.com/books?id=szBLJUhmYOQC}{Applied Combinatorics.}}\/}
	\newblock Boca Raton: CRC Press.
	
	\bibitem[{Rodin et~al.(2011)Rodin, Szathm{\'{a}}ry \& Rodin}]{Rodin2011}
	Rodin, A.~S., Szathm{\'{a}}ry, E. \& Rodin, S.~N. (2011).
	\newblock {\href{http://dx.doi.org/10.1186/1745-6150-6-14}{On origin of genetic code and tRNA before translation.}}
	\newblock {\em Biol. Dir.\/}, {\em 6\/}, 14.
	
	\bibitem[{Ronneberg et~al.(2000)Ronneberg, Landweber \&
		Freeland}]{Ronneberg2000}
	Ronneberg, T.~A., Landweber, L.~F. \& Freeland, S.~J. (2000).
	\newblock {\href{http://dx.doi.org/10.1073/pnas.250403097}{Testing a biosynthetic theory of the genetic code: fact or
			artifact?}}
	\newblock {\em Proc. Natl. Acad. Sci. USA\/},
	{\em 97\/}, 13690--13695.
	
	\bibitem[{Rospars(2010)}]{Rospars2010}
	Rospars, J.-P. (2010).
	\newblock {\href{http://dx.doi.org/10.1016/j.actaastro.2010.03.001}{Terrestrial biological evolution and its implication for SETI.}}
	\newblock {\em Acta Astronaut.\/}, {\em 67\/}, 1361--1365.
	
	\bibitem[{Rumer(1966)}]{Rumer1966}
	Rumer, Y.~B. (1966).
	\newblock {Systematization of codons in the genetic code.}
	\newblock {\em Proceedings of the USSR Academy of Sciences\/}, {\em 167\/},
	1393--1394.
	\newblock (in Russian).
	
	\bibitem[{Rumer(2016)}]{Rumer2016}
	Rumer, Y.~B. (2016).
	\newblock {\href{http://dx.doi.org/10.1098/rsta.2015.0446}{Translation of `Systematization of codons in the genetic code [I]'
			by Yu. B. Rumer (1966).}}
	\newblock {\em Phil. Trans. R. Soc. A\/}, {\em
		374\/}, 20150446.
	
	\bibitem[{Sagan et~al.(1978)Sagan, Drake, Druyan, Ferris, Lomberg \&
		Sagan}]{Sagan1978}
	Sagan, C., Drake, F., Druyan, A., Ferris, T., Lomberg, J. \& Sagan, L.~S.
	(1978).
	\newblock {\em {Murmurs of Earth: The Voyager Interstellar Record.}\/}
	\newblock New York: Random House.
	
	\bibitem[{Sebeok(2001)}]{Sebeok2001}
	Sebeok, T.~A. (2001).
	\newblock {\em {\href{https://books.google.com/books?id=pCbFvfHFitYC}{Signs: An Introduction to Semiotics (2nd ed.).}}\/}
	\newblock Toronto: University of Toronto Press.
	
	\bibitem[{Seife(2000)}]{Seife2000}
	Seife, C. (2000).
	\newblock {\em {\href{https://books.google.com/books?id=0xNvJqQEEvMC}{Zero: The Biography of a Dangerous Idea.}}\/}
	\newblock New York: Penguin Books.
	
	\bibitem[{Shcherbak(1988)}]{Shcherbak1988}
	Shcherbak, V.~I. (1988).
	\newblock {\href{http://dx.doi.org/10.1016/S0022-5193(88)80196-6}{The co-operative symmetry of the genetic code.}}
	\newblock {\em J. Theor. Biol.\/}, {\em 132\/}, 121--124.
	
	\bibitem[{Shcherbak(1989)}]{Shcherbak1989}
	Shcherbak, V.~I. (1989).
	\newblock {\href{http://dx.doi.org/10.1007/BF02388896}{The information artefact of the genetic code.}}
	\newblock {\em Orig. Life Evol. Biosph.\/}, {\em 19\/},
	364--365.
	
	\bibitem[{shCherbak \& Makukov(2013)}]{ShCherbak2013}
	shCherbak, V.~I. \& Makukov, M.~A. (2013).
	\newblock {\href{http://dx.doi.org/10.1016/j.icarus.2013.02.017}{The ``Wow! signal'' of the terrestrial genetic code.}}
	\newblock {\em Icarus\/}, {\em 224\/}, 228--242.
	
	\bibitem[{Shklovskii \& Sagan(1966)}]{Shklovskii1966}
	Shklovskii, I.~S. \& Sagan, C. (1966).
	\newblock {\em {Intelligent Life in the Universe.}\/}
	\newblock San Francisco: Holden-Day.

	\bibitem[{Sleator \& Smith(2017)}]{Sleator2017}
	Sleator, R. \& Smith, N. (2017).
	\newblock {\href{http://dx.doi.org/10.3184/003685017X14901006155062}{Directed panspermia: a 21st century perspective.}}
	\newblock {\em Sci. Prog.\/}, {\em 100(2)\/}, 187--193.
	
	\bibitem[{Sukhotin(1971)}]{Sukhotin1971}
	Sukhotin, B.~V. (1971).
	\newblock {Methods of message decoding}.
	\newblock In S.~A. Kaplan (Ed.) {\em
		\href{https://archive.org/details/nasa_techdoc_19710005278}{Extraterrestrial Civilizations: Problems of Interstellar Communication}\/}, pp. 133--212. NASA Technical Translations, F-631.
	\newblock{(translated from Russian, 1969, Moscow, Nauka.)}
	
	\bibitem[{Swart et~al.(2016)Swart, Serra, Petroni \& Nowacki}]{Swart2016}
	Swart, E.~C., Serra, V., Petroni, G. \& Nowacki, M. (2016).
	\newblock {\href{http://dx.doi.org/10.1016/j.cell.2016.06.020}{Genetic codes with no dedicated stop codon: context-dependent
			translation termination.}}
	\newblock {\em Cell\/}, {\em 166\/}, 691--702.
	
	\bibitem[{Tepfer(2008)}]{Tepfer2008}
	Tepfer, D. (2008).
	\newblock {\href{http://dx.doi.org/10.1016/j.plantsci.2008.08.007}{The origin of life, panspermia and a proposal to seed the Universe.}}
	\newblock {\em Plant Sci.\/}, {\em 175\/}, 756--760.
	
	\bibitem[{Vakoch(1998{\natexlab{a}})}]{Vakoch1998acta}
	Vakoch, D.~A. (1998{\natexlab{a}}).
	\newblock {\href{http://dx.doi.org/10.1016/S0094-5765(98)00029-0}{Constructing messages to extraterrestrials: an exosemiotic
			perspective.}}
	\newblock {\em Acta Astronaut.\/}, {\em 42\/}, 697--704.
	
	\bibitem[{Vakoch(1998{\natexlab{b}})}]{Vakoch1998leo}
	Vakoch, D.~A. (1998{\natexlab{b}}).
	\newblock {\href{http://dx.doi.org/10.2307/1576671}{Signs of life beyond Earth: a semiotic analysis of interstellar
			messages.}}
	\newblock {\em Leonardo\/}, {\em 31\/}, 313--319.
	
	\bibitem[{Vakoch(2011)}]{Vakoch2011}
	Vakoch, D.~A. (2011).
	\newblock {\href{http://dx.doi.org/10.1016/j.actaastro.2010.01.008}{Responsibility, capability, and Active SETI: Policy, law, ethics, and communication with extraterrestrial intelligence.}}
	\newblock {\em Acta Astronaut.\/}, {\em 68\/}, 512--519.
	
	\bibitem[{Vakoch(2014)}]{Vakoch2014}
	Vakoch, D.~A. (Ed.)  (2014).
	\newblock {\em {\href{http://dx.doi.org/10.1007/978-3-642-37750-1}{Extraterrestrial Altruism: Evolution and Ethics in the
				Cosmos.}}\/}
	\newblock Springer.
	
	\bibitem[{Valtonen et~al.(2009)Valtonen, Nurmi, Zheng, Cucinotta, Wilson,
		Horneck, Lindegren, Melosh, Rickman \& Mileikowsky}]{Valtonen2009}
	Valtonen, M., Nurmi, P., Zheng, J.~Q., Cucinotta, F.~A., Wilson, J.~W.,
	Horneck, G., Lindegren, L., Melosh, J., Rickman, H. \& Mileikowsky, C.
	(2009).
	\newblock {\href{http://dx.doi.org/10.1088/0004-637X/690/1/210}{Natural transfer of viable microbes in space from planets in
			extra-solar systems to a planet in our solar system and vice versa.}}
	\newblock {\em Astrophys. J.\/}, {\em 690\/}, 210--215.
	
	\bibitem[{Ward \& Brownlee(2003)}]{Ward2003}
	Ward, P.~D. \& Brownlee, D. (2003).
	\newblock {\em {\href{https://books.google.com/books?id=7HGXqZLmXW4C}{Rare Earth: Why Complex Life Is Uncommon in the Universe.}}\/}
	\newblock New York: Copernicus Books.
	
	\bibitem[{Webb(2015)}]{Webb2015}
	Webb, S. (2015).
	\newblock {\em {\href{http://dx.doi.org/10.1007/978-3-319-13236-5}{If the Universe Is Teeming with Aliens ... Where Is Everybody? Seventy-Five Solutions to the Fermi Paradox and the Problem of Extraterrestrial Life (2nd ed.).}\/}}
	\newblock Springer International Publishing.
	
	\bibitem[{Weiss	 et~al.(2016)Weiss, Sousa, Mrnjavac, Neukirchen, Roettger,
		Nelson-Sathi \& Martin}]{Weiss2016}
	Weiss, M.~C., Sousa, F.~L., Mrnjavac, N., Neukirchen, S., Roettger, M., Nelson-Sathi, S., \& Martin, W.~F. (2016).  
	\newblock {\href{http://dx.doi.org/10.1038/nmicrobiol.2016.116}{The physiology and habitat of the last universal common ancestor.}}
	\newblock {\em Nat. Microbiol.\/}, {\em 1\/}, 16116.
	
	\bibitem[{Wilhelm \& Nikolajewa(2004)}]{Wilhelm2004}
	Wilhelm, T. \& Nikolajewa, S. (2004).
	\newblock {\href{http://dx.doi.org/10.1007/s00239-004-2650-7}{A new classification scheme of the genetic code.}}
	\newblock {\em J. Mol. Evol.\/}, {\em 59\/}, 598--605.
	
	\bibitem[{Wong(1975)}]{Wong1975}
	Wong, J. T.-F. (1975).
	\newblock {\href{http://dx.doi.org/10.1073/pnas.72.5.1909}{A co-evolution theory of the genetic code.}}
	\newblock {\em Proc. Natl. Acad. Sci. USA\/},
	{\em 72\/}, 1909--1912.
	
	\bibitem[{Wong et~al.(2003)}]{Wong2003}
	Wong, P.C., Wong, K., Foote, H. (2003).
	\newblock {\href{http://dx.doi.org/10.1145/602421.602426}{Organic data memory using the DNA approach.}}
	\newblock {\em Commun. ACM\/},
	{\em 46\/}, 95--98.
	
	\bibitem[{Yarus et~al.(2009)Yarus, Widmann \& Knight}]{Yarus2009}
	Yarus, M., Widmann, J.~J. \& Knight, R. (2009).
	\newblock {\href{http://dx.doi.org/10.1007/s00239-009-9270-1}{RNA-amino acid binding: a stereochemical era for the genetic code.}}
	\newblock {\em J. Mol. Evol.\/}, {\em 69\/}, 406--429.
	
	\bibitem[{Yokoo \& Oshima(1979)}]{Yokoo1979}
	Yokoo, H. \& Oshima, T. (1979).
	\newblock {\href{http://dx.doi.org/10.1016/0019-1035(79)90094-0}{Is bacteriophage $\phi$X174 DNA a message from an extraterrestrial
			intelligence?}}
	\newblock {\em Icarus\/}, {\em 38\/}, 148--153.
	
	\bibitem[{Zaitsev(2012)}]{Zaitsev2012}
	Zaitsev, A. (2012).
	\newblock {\href{http://dx.doi.org/10.1016/j.actaastro.2011.05.026}{Classification of interstellar radio messages.}}
	\newblock {\em Acta Astronaut.\/}, {\em 78\/}, 16--19.
	
	\bibitem[{Zhang \& Norman(1995)}]{Zhang1995}
	Zhang, J. \& Norman, D.~A. (1995).
	\newblock {\href{http://dx.doi.org/10.1016/0010-0277(95)00674-3}{A representational analysis of numeration systems.}}
	\newblock {\em Cognition\/}, {\em 57\/}, 271--295.
	
	\bibitem[{Zhang \& Yu(2011)}]{Zhang2011}
	Zhang, Z. \& Yu, J. (2011).
	\newblock {\href{http://dx.doi.org/10.1016/S1672-0229(11)60004-1}{On the organizational dynamics of the genetic code.}}
	\newblock {\em Genomics Proteomics Bioinformatics\/}, {\em 9\/}, 21--29.
	
\end{thebibliography}
\end{document}